\documentclass[letter]{jpsj3}
\usepackage{txfonts}
\usepackage{afterpage}
\usepackage{bm}
\usepackage{amssymb}

\usepackage{color}

\newcommand{\kl}[1]{\left[ #1 \right]}

\newcommand{\ks}[1]{\left( #1 \right)}

\title
{Deformation of the Fermi Surface and Anomalous Mass Renormalization by 
Critical Spin Fluctuations through Asymmetric Spin-Orbit Interaction}
\author
{
Yukinobu {Fujimoto}$^{1}$, Kazumasa {Miyake}$^{2}$, and Hiroyasu {Matsuura}$^{3}$
}

\inst{
$^{1}$Division of Materials Physics, Department of Materials Engineering Science, 
Graduate School of Engineering Science, Osaka University, 
Toyonaka, Osaka 560-8531, Japan\\
$^{2}$Toyota Physical and Chemical Research Institute, Nagakute, Aichi 480-1192, Japan\\
$^{3}$Department of Physics, University of Tokyo, Bunkyo, Tokyo 113-0033, Japan\\
}
\recdate
{
January 21, 2015
}

\abst{
It is shown that an asymmetric spin-orbit interaction is appreciably renormalized by the 
effect of critical spin fluctuations. As an explicit example, the 
Rashba-type interaction is predicted to be suppressed by anti-ferromagnetic critical 
fluctuations near the hot line (or point), leading to a deformation of the Fermi surface near 
the hot line (or point) around which the band 
splitting is minimized, while ferromagnetic critical fluctuations are shown to enhance 
the Rashba-type interaction everywhere on the Fermi surface, leading to an increase 
in the band splitting.  
{
It is also found that the many-body mass renormalization, which diverges towards the magnetic 
quantum critical point, is in the opposite sign for the two split bands owing to the Rashba-type 
interaction.  These predictions are observable through the de Haas-van Alphen experiment in principle.
}
}

\begin{document}
\maketitle
%\parindent=1zw
%\section{Introduction}
The role of asymmetric spin-orbit (ASSO) interaction in giving rise to intriguing 
physical phenomena of strongly correlated electron systems has attracted much attention in 
the past decade.  In particular, it has been discussed in relation to the superconductivity 
of non-centrosymmetric systems  without the inversion centers of  crystals in the field 
of heavy electron systems~\cite{Frigeri1, SFujimoto1}.  It is also an 
essential ingredient for the  surface states of topological insulators, which have been 
extensively discussed for the past several years since their discovery by 
Kane and Mele~\cite{Kane}.  

It is an old problem how many-body effects, owing to the on-site 
Coulomb interaction, affect the atomic spin-orbit (SO) 
interaction in orbitally degenerate $d$-electron systems~\cite{ Yosida,Yafet}.  
Recently, phenomena of {the} growth in the spin Hall effect  have 
attracted much attention, and theoretical proposals have been published from the viewpoint 
of understanding that these phenomena are due to the effect of {the} growth of the atomic 
SO interaction~\cite{Gu1,Gu2}.  

On the other hand, it has also been discussed {how the ASSO interaction of Rashba type 
influences crucially the effect of critical spin fluctuations (CSFs) on the superconducting transition 
temperature near the antiferromagnetic quantum critical point.~\cite{Tada1,Tada2}  
Furthermore, it has recently been discussed} how the Kondo effect is influenced 
if the conduction electrons are subject to strong ASSO interaction.  For example, 
they are concerned with behaviors of magnetic impurities on the surface of topological 
insulators~\cite{Feng, Zitko1} and the Kondo effect in the system where conduction 
electrons are subject to the Rashba-type spin-orbit interaction~\cite{Zitko2,Yanagisawa}.  

However, it has not been clarified how CSFs affect the ASSO interaction.  
This effect possibly gives rise to a deformation of 
the Fermi surface near the magnetic quantum critical point.  
In this Letter, we report on a perturbation calculation of a correction to the ASSO 
interaction under the influence of CSF modes, revealing the fact that CSFs, ferromagnetic (F) 
or antiferromagnetic (AF), give rise to a singular renormalization of the ASSO interaction.  
This fact suggests that the ASSO interaction is renormalized strongly by  many-body effects. 

%\section{Renormalization  of Asymmetric Spin-Orbit Interaction}
%
We discuss the renormalization of the ASSO interaction due to CSFs near the magnetic quantum 
critical point by a perturbation calculation.  Such a renormalization is expected to be visibly 
large considering that CSFs considerably enhance nonmagnetic impurity 
scattering~\cite{Miyake6}.  
We demonstrate that the ASSO interaction is markedly influenced by CSFs, 
{
leading to the deformation of the Fermi surface (FS). We also note that the dynamical effect 
on the renormalization for ASSO interaction gives rise to a nontrivial mass renormalization 
of quasiparticles and the deformation of the FS.}   

{
To provide an explicit discussion, we adopt a single-band model action, 
on the basis of the idea of an itinerant-localized duality model 
with tetragonal symmetry but without an inversion center of the crystal as follows:~\cite{Kuramoto}
\begin{align}
A&=A_{\rm f}+A_{\rm s}+A_{\rm int}+A_{\rm so},
\label{eq:1}
\\
A_{\rm f}&=-\sum_{\mathbf{k},\sigma,n}f_{\mathbf{k}\sigma}^{\,\dagger}(-{i}\varepsilon_{n})
[G_{\sigma}^{(0)}(\mathbf{k},{i}\varepsilon_{n})]^{-1}
f_{\,\mathbf{k}\sigma}({i}\varepsilon_{n}),
\label{eq:2}
\\
A_{\rm s}&=-\sum_{\mathbf{q},m}{\bm S}_{-\mathbf{q}}(-{i}\nu_{m})\cdot{\bm S}_{\mathbf{q}}
({i}\nu_{m})[\chi_{0}({i}\nu_{m})^{-1}-J(\mathbf{q})],
\label{eq:3}
\\
A_{\rm int}&=-\lambda\sum_{\mathbf{k},\mathbf{q},\sigma,\sigma^{\prime}}\sum_{n,m}
f_{-\mathbf{k}-\mathbf{q},\sigma}^{\,\dagger}(-{i}\varepsilon_{n}-{i}\nu_{m})\,
\hat{\bm{\sigma}}_{\sigma,\sigma^{\prime}}
f_{\,\mathbf{k}\sigma^{\prime}}\,({i}\varepsilon_{n})
\cdot{\bm S}_{\mathbf{q}}({i}\nu_{m}),
\label{eq:4}
\\
A_{\rm so}&=\alpha\sum_{\mathbf{k},\sigma,\sigma^{\prime}}f_{\mathbf{k}\sigma}^{\,\dagger}
{\bm{\gamma}}\ks{\mathbf{k}}\cdot\hat{\bm{\sigma}}_{\sigma\sigma^{\prime}}f_{\,\mathbf{k}\sigma^{\prime}},
\label{eq:5}
\end{align}
where $f_{\,\mathbf{k}\sigma}\ks{f_{\mathbf{k}\sigma}^{\,\dagger}}$ is 
the annihilation (creation) operator 
for the quasiparticle, with the momentum $\mathbf{k}$ and the ``spin" $\sigma$, renormalized 
by local correlation, 
and ${\bm S}\,(\mathbf{q})$ is the local spin density operator constructed from high-energy states of 
correlated electrons.  Note that ``spin" stands for pseudo-spin specifying a degeneracy of the Kramers 
doublet in general.  Hereafter, we call it as simply spin.  
The Green function $G_{\sigma}^{(0)}(\mathbf{k},{i}\varepsilon_{n})$ of 
the quasiparticle is given by 
\begin{align}
G_{\sigma}^{(0)}(\mathbf{k},{i}\varepsilon_{n})\simeq
\frac{z}{{i}\varepsilon_{n}-\xi_{\mathbf{k}}},
\label{rashba7}
\end{align}
where  $\xi_\mathbf{k}$ is the dispersion of the quasiparticle measured from the Fermi energy, 
$\varepsilon_n=(2n+1)\pi T$ is the fermionic Matsubara frequency, 
and $z$ is the renormalization amplitude due to the effect of the local correlation.  
$\chi_{0}({i}\nu_{m})$ (with $\nu_m=2m\pi T$ being the bosonic Matsubara frequency) 
and $J(\mathbf{q})$ in Eq.\ (\ref{eq:3}) represent the local dynamical spin 
susceptibility and the exchange interaction among localized parts of spin degrees of 
freedom, respectively.  
$A_{\rm int}$ [Eq.\ (\ref{eq:4})] represents the exchange interaction between quasiparticles and 
local spins.  Finally, $A_{\rm{so}}$ [Eq.\ (\ref{eq:5})] represents the ASSO 
interaction:~\cite{Frigeri1, SFujimoto1}   
$\alpha$ is the coupling constant of the ASSO interaction, and 
$\mathbf{\hat{\bm{\sigma}}}=\ks{\sigma_x,\sigma_y,\sigma_z}$ is the} 
{vector spanned by the Pauli matrices}.  {
The vector ${\bm{\gamma}}\ks{\mathbf{k}}=(\gamma_{k_x},\gamma_{k_y},\gamma_{k_z})$ in Eq.\ (\ref{eq:5}) 
obeys ${\bm{\gamma}}\ks{-\mathbf{k}}=-{\bm{\gamma}}\ks{\mathbf{k}}$, thereby breaking the inversion 
symmetry and causing the ASSO interaction.}  {The main purpose of this study 
is to demonstrate 
how the ASSO coupling constant $\alpha$ is renormalized by CSFs.}  { 
Hereafter, to provide explicit discussions, 
we assume that the ASSO interaction is the Rashba-type SO interaction given by~\cite{Rashba} 
\begin{align}
{\bm{\gamma}}\ks{\mathbf{k}}=\ks{\sin k_{y}a,-\sin k_{x}a,0}, 
\label{rashba4.3}
\end{align}
where $a$ is the lattice constant in the tetragonal plane.  
However, it remains valid also for another type of ASSO interaction~\cite{Dresselhaus,Samokhin} 
that the ASSO coupling constant $\alpha$ is considerably influenced by CSFs.  

The lowest-order correction to the coupling constant $\alpha$ of the ASSO interaction 
due to CSFs is depicted by the Feynman diagrams shown in Fig.\ \ref{htm01}.  
The off-diagonal (in spin space) self-energy of quasiparticles due to this correction 
$\Sigma_{{\rm{I\hspace{-.1em}I}}\uparrow\downarrow}\ks{\mathbf{k},\rm{i}\varepsilon_{n}}$ 
appearing in Fig.\ \ref{htm01}(a) is given by 
\begin{align}
%\sum
\Sigma_{{\rm{I\hspace{-.1em}I}}\uparrow\downarrow}\ks{\mathbf{k},{i}\varepsilon_n}
=
\lambda^2 T\sum\limits_{\mathbf{k}^\prime}\sum\limits_{n^{\,\prime}}\alpha
[{\bm{\gamma}}\ks{\mathbf{k}^\prime}\cdot\mathbf{\hat{\bm{\sigma}}}_{\uparrow\downarrow}]\,
\chi_{\uparrow\downarrow}\ks{\mathbf{k}-\mathbf{k}^\prime,{i}\varepsilon_{n}
-{i}\varepsilon_{n^{\prime}}}
[G^{(0)}\ks{\mathbf{k}^\prime,{i}\varepsilon_{n^{\prime}}}]^{\,2},
\label{rashba4}
\end{align}
where $\chi_{\uparrow\downarrow}$ is a part of the spin-fluctuation propagator 
$\chi_{\rm s}=(\chi_{\uparrow\uparrow}-\chi_{\uparrow\downarrow})/2$.  

\begin{figure}
[!htb]
%[htbp]
\begin
{center}
\includegraphics [scale=0.7]{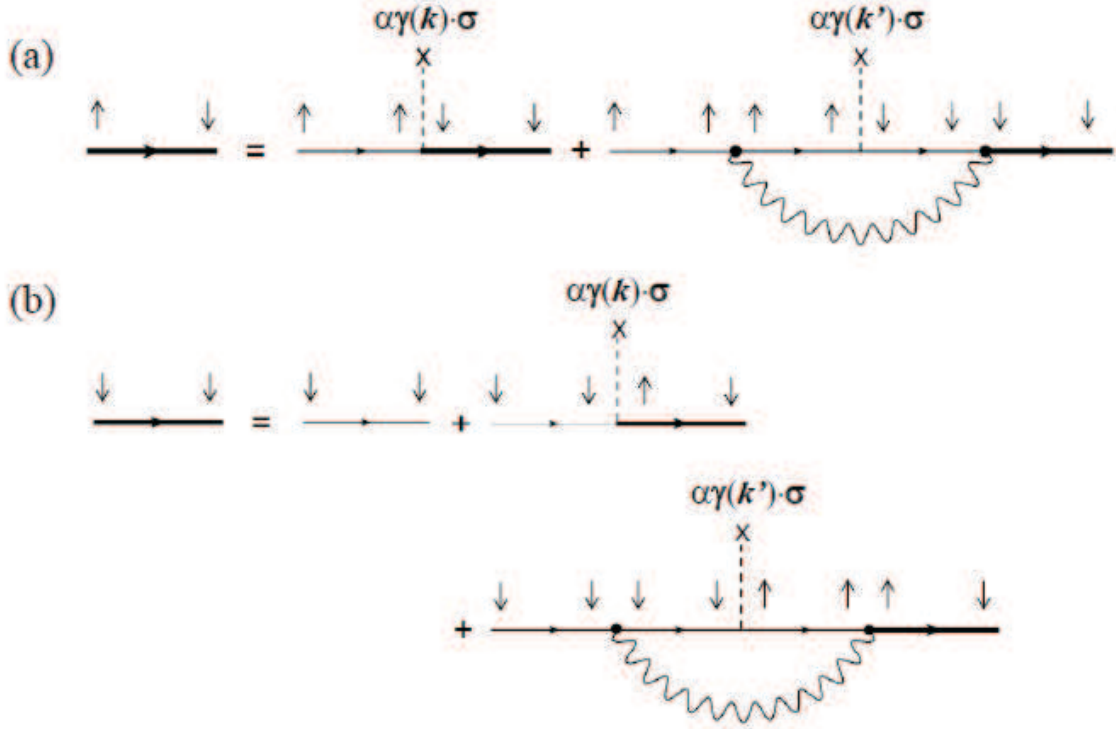}
\caption{Feynman diagrams for the lowest-order correction to the coupling constant $\alpha$ of 
the ASSO interaction due to CSFs. The wavy line represents the spin-fluctuation propagator 
$\chi_s$, the solid line the Green function $G$ of the quasiparticles, the filled circle the coupling 
constant $\lambda$ between spin fluctuations and quasiparticles, the upward arrow the up 
spin, the downward arrow the down spin, and the cross the ASSO interaction.} 
\label{htm01}
\end{center}
\end{figure}
%
%
%%%%%%%%%%%%%%%%%%%%%
%
%%%%%%%%%%%%%%%%%%%

The propagator $\chi_{\uparrow\downarrow}$ is expressed in terms of $\chi_{\rm s}$ and 
the charge-fluctuation propagator 
$\chi_{\rm c}=2(\chi_{\uparrow\uparrow}+\chi_{\uparrow\downarrow})$ as 
$\chi_{\uparrow\downarrow}=-\chi_{\rm s}+(\chi_{c}/4)${, in general}.  
Near the magnetic critical point, 
the propagator $\chi_{\uparrow\downarrow}$ is dominated by the spin-fluctuation propagator 
$\chi_{\rm s}$ and is given by~\cite{Miyake6,Moriya1,Fujimoto2, Fujimoto3}  
\begin{align}
\chi_{\uparrow\downarrow}\ks{\mathbf{q},{i}\omega_{m}}
\simeq
-\frac{\chi^{\ks{0}}_\mathbf{Q}}{\eta + A\ks{\mathbf{q}-\mathbf{Q}}^2+C_q|\omega_{m}|},
\label{rashba7.4}
\end{align}
where $\omega_m=2m\,\pi T$ is the boson Matsubara frequency, $\eta$ is the distance from the 
magnetic critical point, $\mathbf{Q}$ is the magnetic ordering vector, and 
$\chi^{\ks{0}}_\mathbf{Q}$ is the noninteracting static spin susceptibility at the 
magnetic ordering vector $\mathbf{Q}$.  Note that $\eta>0$ represents the paramagnetic state 
and $\eta=0$ corresponds to the  magnetic critical point~\cite{ Fujimoto2, Fujimoto3}. 

Substituting Eq.\ (\ref{rashba7.4}) into Eq.\ (\ref{rashba4}), we obtain
\begin{align}
\Sigma_{{\rm{I\hspace{-.1em}I}}\uparrow\downarrow}
\ks{\mathbf{k},{i}\varepsilon_{n}}
&
=\sum_{\nu=x,y}(\sigma_{\nu})_{\uparrow\downarrow}\Gamma_{\nu}\ks{\mathbf{k},{i}\varepsilon_{n}},
\label{rashba4.18}
\end{align}
with
\begin{align}
\Gamma_{\nu}\ks{\mathbf{k},{i}\varepsilon_{n}}
&
\equiv -\alpha
{\tilde \lambda}^{2}T\sum\limits_{\mathbf{k}^\prime}\sum\limits_{n^{\,\prime}}
\gamma_{\nu}(\mathbf{k}^\prime)
\frac{\chi_{\bf Q}^{(0)}}{\eta+A({\bf k}-{\bf k}^{\prime}-{\bf Q})^{2}
+C_{|{\bf k}-{\bf k}^{\prime}|}\,|\varepsilon_{n}-\varepsilon_{n^{\,\prime}}|}
\frac{1}{({i}\varepsilon_{n^{\,\prime}}-{\xi}_{{\bf k}^{\prime}})^{2}}
\quad
\ks{\nu=x,y},
\label{rashba4.4}
\end{align}
where ${\tilde \lambda}\equiv {z}\lambda$.  The other off-diagonal self-energy 
$\Sigma_{{\rm{I\hspace{-.1em}I}}\downarrow\uparrow}
\ks{\mathbf{k},{i}\varepsilon_{n}}$, shown in Fig.\ \ref{htm01}(b), is given by
\begin{align}
\Sigma_{{\rm{I\hspace{-.1em}I}}\downarrow\uparrow}
\ks{\mathbf{k},{i}\varepsilon_{n}}
&
=\sum_{\nu=x,y}(\sigma_{\nu})^{*}_{\uparrow\downarrow}\Gamma^{*}_{\nu}\ks{\mathbf{k},
-{i}\varepsilon_{n}},
\label{rashba4.18a}
\end{align}
where we have used a general relation
\begin{equation}
G_{\downarrow\uparrow}(\mathbf{k},{i}\varepsilon_{n})=
[G_{\uparrow\downarrow}(\mathbf{k},-{i}\varepsilon_{n})]^{*},
\label{rashba4.18b}
\end{equation}
as shown in Sect.\ 1 of Ref.\ \citen{Suppl}.

First, we discuss the effect of three-dimensional (3D) 
AF quantum critical fluctuations (QCF{s}) in which 
$C_{q}$ is $q$-independent, i.e., $C_{q}\equiv C$.  
{
We also discuss the case where the external wave vector ${\bf k}$ is on the FS.
}
To begin with, let us define ${\bf k}^{*}$ as 
${\bf k}^{*}\equiv {\bf k}-{\bf Q}$ 
{
and investigate the case where ${\bf k}^{*}$ 
is on the spherical FS; i.e., {\bf k} is on the hot line where 
}
$\xi_{{\bf k}+{\bf Q}}=-{\xi}_{\bf k}$.  
Note that the results {obtained in the present study} are valid for any shape of FS 
as long as Eq.\ (\ref{rashba7.4}) is valid, 
{
although the calculations become far more tedious. }  
Here we assume that the origin of the 
AF order is caused by a commensurate AF-exchange interaction through an incoherent process 
in the meaning of the itinerant-localized duality model 
{
so that the ordering vector ${\bf Q}$ is not influenced by QCFs
.~\cite{Kuramoto,Miyake2}}  

Then, the frequency-independent terms in the denominator of Eq.\ (\ref{rashba7.4}) are transformed into 
\begin{equation}
\eta+A({\bf k}-{\bf k}^{\prime}-{\bf Q})^{2}
=\eta+A({\bf k}^{*}-{\bf k}^{\prime})^{2}
=\frac{A}{v_{\rm F}^{\,2}}(\xi^{\prime 2}+2b\,\xi^{\prime}+a),
\label{eq:M1}
\end{equation}
where $\xi^{\prime}$ stands for ${\xi}_{{\bf k}^{\prime}}$, and $a$ and $b$ are 
defined by 
\begin{equation}
a\equiv \frac{v_{\rm F}^{\,2}}{A}\eta+2k_{\rm F}^{\,2}v_{\rm F}^{\,2}(1-\cos\,\theta),
\label{eq:M2}
\end{equation}
and 
\begin{equation}
b\equiv k_{\rm F}v_{\rm F}(1-\cos\,\theta),
\label{eq:M3}
\end{equation}
where $\cos\,\theta\equiv({\bf k}^{*}\cdot{\bf k}^{\prime})/(|{\bf k}^{*}||{\bf k}^{\prime}|)$, 
and $k_{\rm F}$ and $v_{\rm F}$ are the Fermi wavenumber and velocity of quasiparticles on the 
FS, respectively. Thus, $\Gamma_{\nu}({\bf k},{i}\varepsilon_{n})$ [Eq.\ (\ref{rashba4.4})] 
is expressed as 
\begin{equation}
\Gamma_{\nu}\ks{\mathbf{k},{i}\varepsilon_{n}}
= -\alpha{\tilde \lambda}^{2}\gamma_{\nu}({\bf k}-{\bf Q})\,
T\sum_{n^{\,\prime}}\int_{-1}^{1}\frac{{d}(\cos\theta)}{2}\int{d}\xi^{\prime}N(\xi^{\prime})
\frac{v_{\rm F}^{\,2}}{A}\frac{\chi_{Q}^{(0)}}
{\xi^{\prime\,2}+2b\,\xi^{\prime}+a+C^{*}|\varepsilon_{n}-\varepsilon_{n^{\,\prime}}|}
\frac{1}{(\xi^{\prime}-{i}\varepsilon_{n^{\,\prime}})^{2}},
\label{eq:M4}
\end{equation}
where $C^{*}\equiv v_{\rm F}^{\,2}C/A$.  
Since the singular behavior in $\Gamma_{\nu}$ appears through the contributions from the regions 
$\xi^{\prime}\simeq 0$ and $\cos\,\theta\simeq 1$, in deriving Eq.\ (\ref{eq:M4}), 
we have used the approximation ${\bf k}^{\prime}\approx {\bf k}^{*}={\bf k}-{\bf Q}$; 
i.e., $\gamma_{\nu}({\bf k}^{\prime})\approx\gamma_{\nu}({\bf k}-{\bf Q})$.  
Similarly, the density of states of quasiparticles, $N(\xi^{\prime})$, is approximated by 
that at the Fermi level, $N_{\rm F}$.  Therefore, $\Gamma_{\nu}({\bf k},{i}\varepsilon_{n})$ 
is reduced to 
\begin{equation}
\Gamma_{\nu}\ks{\mathbf{k},{i}\varepsilon_{n}}
= -J\gamma_{\nu}({\bf k}-{\bf Q})\,
\int_{-1}^{1}\frac{{d}(\cos\theta)}{2}T\sum_{n^{\,\prime}}\int{d}\xi^{\prime}
\frac{1}{\xi^{\prime\,2}+2b\,\xi^{\prime}+a+C^{*}|\varepsilon_{n}-\varepsilon_{n^{\,\prime}}|}
\frac{1}{(\xi^{\prime}-{i}\varepsilon_{n^{\,\prime}})^{2}}, 
\label{eq:M5}
\end{equation}
where $J$ is defined by 
\begin{equation}
J\equiv\alpha{\tilde \lambda}^{2}N_{\rm F}\chi_{Q}^{(0)}\frac{v_{\rm F}^{\,2}}{A}. 
\label{eq:M6}
\end{equation}

Performing the integration with respect to $\xi^{\prime}$ in Eq.\ (\ref{eq:M5}) with the use of 
the contour integration, we obtain 
\begin{eqnarray}
& &\Gamma_{\nu}\ks{\mathbf{k},{i}\varepsilon_{n}}
= -J\gamma_{\nu}({\bf k}-{\bf Q})\,\int_{-1}^{1}\frac{{d}(\cos\theta)}{2}
\nonumber
\\
& &\quad
\pi 
T\sum_{n^{\,\prime}\ge 0}\left\{
\frac{1}{\sqrt{{\tilde a}_{+}-b^{\,2}}}
\frac{1}{[b+{i}(\sqrt{{\tilde a}_{+}-b^{\,2}}+\varepsilon_{n^{\,\prime}})]^{2}}
+
\frac{1}{\sqrt{{\tilde a}_{-}-b^{\,2}}}
\frac{1}{[b-{i}(\sqrt{{\tilde a}_{-}-b^{\,2}}+\varepsilon_{n^{\,\prime}})]^{2}}
\right\}, 
\label{eq:M7}
\end{eqnarray}
where ${\tilde a}_{\pm}\equiv a+C^{*}|\pm\varepsilon_{n}-\varepsilon_{n^{\,\prime}}|$.
{  
Here we note that Eq.\ (\ref{eq:M7}) is valid regardless of the sign of $\varepsilon_{n}$, 
positive or negative, although we consider the case with $\varepsilon_{n}>0$ hereafter.  
Equation (\ref{eq:M7}) is rearranged and approximated} as 
\begin{eqnarray}
& &
\Gamma_{\nu}\ks{\mathbf{k},{i}\varepsilon_{n}}=
{\bar \Gamma}_{\nu}\ks{\mathbf{k},{i}\varepsilon_{n}}+
J\gamma_{\nu}({\bf k}-{\bf Q})\,\int_{-1}^{1}\frac{{d}(\cos\theta)}{2}
\nonumber
\\
& &
\qquad\qquad\qquad\qquad
{\pi T}
\left\{
{
\sum_{n^{\,\prime}\ge 0,\,n^{\,\prime}\not=n}
}
\left[
\frac{1}{\sqrt{{\tilde a}_{+}}}\frac{1}{(\sqrt{{\tilde a}_{+}}+\varepsilon_{n^{\,\prime}})^{2}}+
\frac{1}{\sqrt{{\tilde a}_{-}}}\frac{1}{(\sqrt{{\tilde a}_{-}}+\varepsilon_{n^{\,\prime}})^{2}}
\right]
\nonumber
\right.
\\
& &
\left.
\qquad\qquad\qquad\qquad\quad
+2{i}b{\sum_{n^{\,\prime}\ge 0}}
\left[
\frac{1}{\sqrt{{\tilde a}_{+}}}\frac{1}{(\sqrt{{\tilde a}_{+}}+\varepsilon_{n^{\,\prime}})^{3}}-
\frac{1}{\sqrt{{\tilde a}_{-}}}\frac{1}{(\sqrt{{\tilde a}_{-}}+\varepsilon_{n^{\,\prime}})^{3}}
\right]
\right\},
\label{eq:M8}
\end{eqnarray}
where ${\bar \Gamma}_{\nu}\ks{\mathbf{k},{i}\varepsilon_{n}}$ is the contribution from 
$n^{\,\prime}=n>0$ 
{
in the first term in the brace of Eq.\ (\ref{eq:M7}), 
} 
{which corresponds to the static term of fluctuation and should be treated 
separately, as discussed in Sect.\ 3 of Ref.\ \citen{Suppl}.}  In deriving Eq.\ (\ref{eq:M8}), 
we have neglected the terms of ${\cal O}(b^{2})$ because the singular contribution 
to $\Gamma_{\nu}$ arises from the regions $\cos\,\theta\simeq 1$ and 
$b^{\,2}\propto (1-\cos\,\theta)^{2}$.  Then, the retarded function 
$\Gamma_{\nu}^{\rm R}\ks{\mathbf{k},\varepsilon+{i}\delta}$ at $T\to 0$ in the static limit 
($\varepsilon\to 0$) is 
given by $\lim_{T\to 0}\Gamma_{\nu}\ks{\mathbf{k},{i}\pi T}$:
\begin{equation}
\Gamma_{\nu}^{\rm R}\ks{\mathbf{k},{i}\delta}
= {{\bar \Gamma}_{\nu}^{\rm R}\ks{\mathbf{k},{i}\delta}+}
J\gamma_{\nu}({\bf k}-{\bf Q})\,\int_{-1}^{1}\frac{{d}(\cos\theta)}{2}
\int_{0}^{\infty}{d}\varepsilon^{\prime}
\frac{1}{\sqrt{a+C^{*}\varepsilon^{\prime}}
(\sqrt{a+C^{*}\varepsilon^{\prime}}+\varepsilon^{\prime})^{2}}, 
\label{eq:M9}
\end{equation} 
where { the term ${\bar \Gamma}^{\rm R}_{\nu}\ks{\mathbf{k},{i}\delta}$ is given 
by the first term in the bracket of Eq.\ (23)} in Sect.\ 3 of Ref.\ \citen{Suppl}. 
By changing the integration variable from $\varepsilon^{\prime}$ to 
$y\equiv \sqrt{a+C^{*}\varepsilon^{\prime}}/\sqrt{a}$, Eq.\ (\ref{eq:M9}) is reduced to a 
more compact form as 
\begin{equation}
\Gamma_{\nu}^{\rm R}\ks{\mathbf{k},{i}\delta}
= \alpha_{\rm af}{(\eta)}\gamma_{\nu}({\bf k}-{\bf Q}),
\label{eq:M10z}
\end{equation}
where the excess contribution $\alpha_{\rm af}$ to the SO coupling from the spin fluctuations 
is given by 
\begin{equation}
\alpha_{\rm af}{(\eta)=\frac{{\bar \Gamma}_{\nu}^{\rm R}\ks{\mathbf{k},{i}\delta}}
{\gamma_{\nu}({\bf k}-{\bf Q})}+}
\frac{2J}{C^{*}}\int_{-1}^{1}\frac{{d}(\cos\theta)}{2}
\frac{1}{\sqrt{a}}\int_{1}^{\infty}{d}y
\frac{1}{\left[y+\frac{\sqrt{a}}{C^{*}}(y^{\,2}-1)\right]^{2}}.
\label{eq:M10}
\end{equation} 
With the use of the expression for $a$ [Eq.\ (\ref{eq:M2})], it is easier to perform first 
the integration with respect to $\cos\theta$ and then with respect to $y$.  
{According to Sects.\ 2 and 3 of Ref.\ \citen{Suppl}, the result for 
$\alpha_{\rm af}$ up to ${\cal O}(\sqrt{\eta})$ is given by}  
{
\begin{equation}
\alpha_{\rm af}{(\eta)}\simeq 
\frac{J}{2k_{\rm F}^{\,2}v_{\rm F}^{\,2}}\left[\,F\left(\frac{{4}m^{*}A}{C}\right)
-\frac{2\sqrt{A}}{v_{\rm F}C}\sqrt{\eta}+
{\frac{\sqrt{Ak_{\rm F}^{2}}}{k_{\rm F}v_{\rm F}}
\frac{\pi T}{\sqrt{\eta}}+}
{\cal O}(\eta)\right],
\label{eq:M11}
\end{equation} 
where $m^{*}$ is the effective mass of quasiparticles and $F(x)$ is defined by 
\begin{equation}
F(x)\equiv \log\frac{x}{2}
-\frac{1}{\sqrt{1+x^{\,2}}}\log\frac{x+1-\sqrt{1+x^{\,2}}}{x+1+\sqrt{1+x^{\,2}}}. 
\label{eq:M12}
\end{equation}
The function $F(x)$ is an increasing function and positive definite for $x>0$, 
and $F(1)\simeq 0.5{5}$.  Therefore, $\Gamma_{\nu}^{\rm R}\ks{\mathbf{k},{i}\delta}$ 
has a cusp singularity at $\eta=0$.  

The $\varepsilon_{n}$ dependence of $\Gamma_{\nu}\ks{\mathbf{k},{i}\varepsilon_{n}}$ for 
$\varepsilon_{n}>0$, 
up to ${\cal O}(\varepsilon_{n})$, is given in Sect.\ 3 of Ref.\ \citen{Suppl} as 
\begin{equation}
\Gamma_{\nu}\ks{\mathbf{k},{i}\varepsilon_{n}}=
\Gamma_{\nu}^{\rm R}\ks{\mathbf{k},{i}\delta}+
\gamma_{\nu}({\bf k}-{\bf Q})\,h\,({i}\varepsilon_{n})
+{\cal O}(\varepsilon_{n}^{2}),
\label{eq:M9a}
\end{equation}
where $h$ is defined by 
}
{
\begin{equation}
h\equiv J\left[-\frac{\sqrt{Ak_{\rm F}^{2}}}{(k_{\rm F}v_{\rm F})^{4}}\,
\frac{\pi T}{\sqrt{\eta}}
+\frac{1}{4(k_{\rm F}v_{\rm F})^{3}}\left(\log\,\frac{4Ak_{\rm F}^{2}}{\eta e}\right)\right].
\label{eq:M9b2}
\end{equation}
}
{
We note that $h$ is divergent because $\eta\propto T^{3/2}$ just at the 
3D-AFQCP.~\cite{MoriyaTakimoto}}  {In the case where the result of the first-order 
perturbation calculation is divergent,} {we need to consider the effect of 
higher-order corrections with respect to 
CSFs in general. This problem is beyond our scope and will be left as a future work. 
}

Substituting Eq.\ (\ref{eq:M9a}) into Eqs.\ (\ref{rashba4.18}) and (\ref{rashba4.18a}) and performing 
the analytic continuation ${i}\varepsilon_{n}\to \varepsilon+{i}\delta$, 
we obtain
{
\begin{equation}
\Sigma_{\rm{I\hspace{-.1em}I}\uparrow\downarrow}^{\rm R}
\ks{\mathbf{k},\varepsilon+{i}\delta}
\simeq
{\bm{\gamma}}\ks{\mathbf{k}-\mathbf{Q}}\cdot(\mathbf{\hat{\bm{\sigma}}})_{\uparrow\downarrow}
[\alpha_{\rm {af}}\ks{\eta}+h\,\varepsilon],
\label{rashba4.13a}
\end{equation}
and 
\begin{equation}
\Sigma_{\rm{I\hspace{-.1em}I}\downarrow\uparrow}^{\rm R}
\ks{\mathbf{k},\varepsilon+{i}\delta}
\simeq
{\bm{\gamma}}\ks{\mathbf{k}-\mathbf{Q}}\cdot(\mathbf{\hat{\bm{\sigma}}})^{*}_{\uparrow\downarrow}
[\alpha_{\rm {af}}\ks{\eta}+h\,\varepsilon].
\label{rashba4.13b}
\end{equation}
Then, the retarded Green function matrix $\hat{G}^{\rm R}$ for quasiparticles, renormalized by QCFs, 
is given by
\begin{align}
\hat{G}^{\rm R}(\mathbf{k},\varepsilon)
=
\kl{\ks{\varepsilon-\xi_{\mathbf{k}}}\hat{1}
-
\alpha{\bm{\gamma}}\ks{\mathbf{k}}\cdot\mathbf{\hat{\bm{\sigma}}}
-
\hat{\Sigma}^{\rm R}\ks{\mathbf{k},\varepsilon}}^{-1},
\end{align}
with
\begin{align}
\hat{\Sigma}^{\rm R}\ks{\mathbf{k},\varepsilon}
=
\Sigma_{\rm{I}}^{\rm R}\ks{\mathbf{k},\varepsilon}\hat{1}
+
\hat{\Sigma}_{\rm{I\hspace{-.1em}I}}^{\rm R}\ks{\mathbf{k},\varepsilon}.
\label{rashba4.9}
\end{align}
Here, $\Sigma_{\rm{I}}^{\rm R}\ks{\mathbf{k},\varepsilon}$ is 
the self-energy of the quasiparticles due to 
the effect of spin fluctuations without considering the Rashba-type SO interaction.
}

According to Sect.\ 4 of Ref.\ \citen{Suppl}, by diagonalizing the Green function matrix 
$\hat{G}^{\rm R}$, {the dispersion of quasiparticles is given by 
\begin{equation}
\tilde{\xi}_{\mathbf{k}}^{\pm}
=\left[\xi_{\mathbf{k}}+\Sigma_{\rm I}^{\rm R}(\mathbf{k},0)
\pm\left |\alpha{\bm{\gamma}}\ks{\mathbf{k}}
+\alpha_{\rm af}(\eta){\bm{\gamma}}\ks{\mathbf{k}-\mathbf{Q}}\right|\right]{\tilde z}^{\pm}_{\mathbf{k}},
\label{rashba8.3}
\end{equation}
where}
{
\begin{equation}
\left({\tilde z}^{\pm}_{\mathbf{k}}\right)^{-1}
=1-
\left.\frac{\partial \Sigma_{\rm{I}}^{\rm R}\ks{\mathbf{k},\varepsilon}}
{\partial \varepsilon}\right|_{\varepsilon=0}
\mp\frac{h\,[\alpha\bm{\gamma}({\bf k})+\alpha_{\rm af}(\eta)\bm{\gamma}({\bf k}-{\bf Q})]
\cdot\bm{\gamma}({\bf k}-{\bf Q})}{|\alpha\bm{\gamma}({\bf k})+\alpha_{\rm af}(\eta)
\bm{\gamma}({\bf k}-{\bf Q})|}.
\label{rashba11.2AF}
\end{equation}
}
It can be seen from Eqs.\ (\ref{rashba8.3}) and (\ref{rashba11.2AF}) that the band (0) with the energy 
{
$\tilde{\xi}_{\mathbf{k}}^0=[\xi_{\mathbf{k}}
+\Sigma_{\rm I}^{\rm R}(\mathbf{k},0)]{\tilde z}_{\mathbf{k}}$, where 
${\tilde z}_{\mathbf{k}}$ is equal to ${\tilde z}^{\pm}_{\mathbf{k}}$ with $h=0$, 
} is split into the two bands ($\pm$) with the energies 
$\tilde{\xi}_{\mathbf{k}}^{\pm}$, and the term of $\alpha_s(\eta){\bm{\gamma}}\ks{\mathbf{k}-\mathbf{Q}}$ 
is newly involved into the coupling constant of the Rashba-type SO interaction due to CSFs. 
{
The FS is deformed by an effect of $\Sigma_{\rm I}(\mathbf{k},0)$ in general.  
Therefore, precisely speaking, the calculation of 
${\hat \Sigma}_{\rm II}^{\rm R}(\mathbf{k},\varepsilon)$ in Eq.\ (\ref{rashba4.9}) 
should have been performed 
with the use of quasiparticles with this deformed FS.  However, we are interested in the effect of 
${\hat \Sigma}_{\rm II}^{\rm R}(\mathbf{k},\varepsilon)$ on the band splitting due to the 
ASSO interaction, and 
we have taken an approximation that the spin-fluctuation propagator (\ref{rashba7.4}) 
is maintained intact.  
}
{Note} {also} {that the divergences in  
$h$ [Eq.\ (\ref{eq:M9b2})] can make either of $\left({\tilde z}^{\pm}_{\mathbf{k}}\right)^{-1}$ 
negative, which is unphysical.  To remedy this fault, we need to consider the effect of 
higher-order corrections with respect to CSFs, as pointed out above.
}

{
The derivations above are restricted to the case where ${\bf k}$ is on the hot line(s) on the FS  
without the ASSO interaction.   
To discuss a deformation of the {FS} due to the ASSO interaction renormalized by 
the AF-CSF, {the effect of} the distance of the position of ${\bf k}$ from the hot line on 
the FS {is parameterized} by shifting $\eta$ to $\eta+Aq_{m}^{\,2}$ 
in Eq.\ (\ref{eq:M11}){,}~\cite{Miyake6}  
{where $q_{m}$ is the perpendicular distance between ${\bf k}$ and the hot line on the FS 
without the ASSO interaction.} 
Then, 
{
in the case where the ordering vector is commensurate, i.e., ${\bf Q}=(\pi/a,\pi/a,\pi/c)$, 
so that $\gamma_{\nu}({\bf k}-{\bf Q})=-\gamma_{\nu}({\bf k})$,  
}
the band energy $\tilde{\xi}_{\mathbf{k}}^{\pm}$ in Eq.\ (\ref{rashba8.3}) is approximated as 
{
\begin{align}
\tilde{\xi}_{\mathbf{k}}^{\pm}
&\simeq
\left[\xi_{\mathbf{k}}+\Sigma_{\rm I}^{\rm R}(\mathbf{k},0)
\pm
\left|\alpha-\alpha_{\rm af}\left(\eta+Aq_{m}^{\,2}\right)\right|
\left|{\bm{\gamma}}\ks{\mathbf{k}}\right|
\right]{\tilde z}^{\pm}_{\mathbf{k}},
\label{rashba11.2b}
\end{align}
}
where ${\tilde z}^{\pm}_{\mathbf{k}}$ is given by modifying $h$ in Eq.\ (\ref{rashba11.2AF}) with 
the replacement of $\eta$ by $\eta+Aq_{m}^{2}$ in Eq.\ (\ref{eq:M9b2}).  
Therefore, {according to Eq.\ (\ref{rashba11.2AF}),} 
we expect that the effect of dynamical mass renormalization 
is opposite in the two split bands: 
the effective mass in band {(+)} is {enhanced} while that 
in band {(-)} is {suppressed}. 

The singular behavior given by Eq.\ (\ref{eq:M11}) fades away at positions on the {FS} 
far from the hot lines.  
As a result, the suppression of the band splitting is confined to the region near the hot line, 
as shown in  Fig. \ref{rashba9.0}(b). Thus, it means that the deformation of the {FS} 
is caused by the band splitting under the influence of  critical AF spin fluctuations.  
}
{In particular, the suppression of the band splitting seems to be followed by an AF-gap 
opening around the hot line(s), which is reasonable from the physical viewpoint.  }
{In the ``AF ordered state", the ordering vector $\bf Q$ itself should be determined 
self-consistently by taking into account the effect of the enhancement of the ASSO interaction $\alpha$.  
Therefore, it is possible that a resultant ordering vector ${\bf Q}$ can be incommensurate 
even though the critical mode is commensurate in the paramagnetic state.  
This problem will be discussed elsewhere.  }

%
%%%%%%%%%%%%%%%%%%%
%
\begin{figure}
[!htb]
%[htbp]
\begin
{center}
\includegraphics [scale=0.3]
{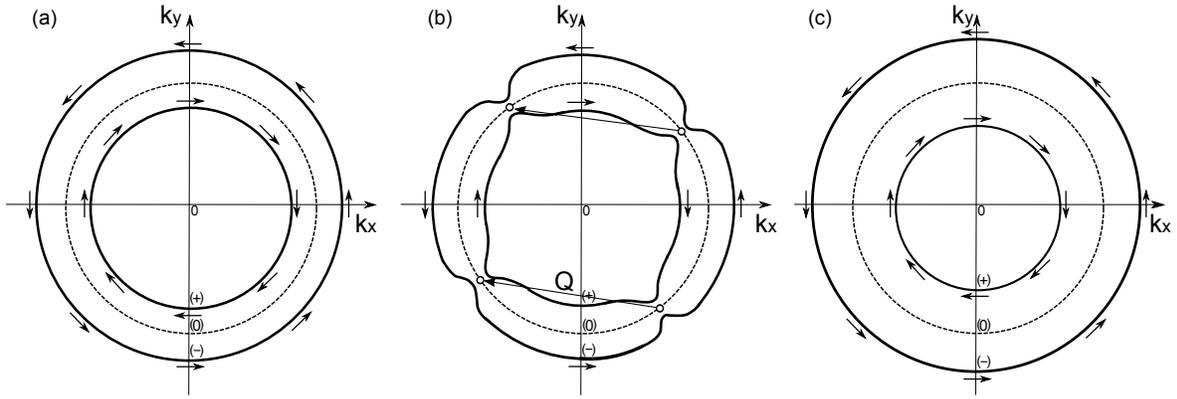}
\caption{Fermi surfaces split by the Rashba-SO interaction for (a) the system without 
{the effect of CSFs}, 
(b) the system with critical AF spin fluctuations, and (c)  the system with 
critical F spin fluctuations. The circle indicated by the dotted line  represents the  Fermi surface of 
the band (0), the large circle indicated by the solid line with the ($\pm$) represents the 
Fermi surface of 
the band $(\pm)$, the long leftward arrow represents the AF-order vector $\mathbf{Q}$,  
the short arrow represents the direction of the spin,  and the small open circle indicated by 
the solid line represents the hot spot.}
\label{rashba9.0}
\end{center}
\end{figure}
%%%%%%%%%%%%%%%%%%%%%%%

{
There exists a singular contribution to the renormalization amplitude 
${\tilde z}^{\pm}_{\mathbf{k}}$ 
near the criticality ($\eta=0$) through a singularity included in $h$ [Eq.\ (\ref{eq:M9b2})].  
Another nontrivial aspect suggested by Eqs.\ (\ref{rashba8.3}) and (\ref{rashba11.2AF}) is that 
the mass renormalization effects [through the factor $\left({\tilde z}^{\pm}_{\mathbf{k}}\right)^{-1}$] 
for the two split bands are in the opposite sign, as 
explicitly shown in Eq.\ (\ref{rashba11.2AF}).  This effect can be observed through 
the de Haas-van Alphen {(dHvA)} experiment in principle.  

Note that the aspects discussed above are expected to survive even if the ordering vector 
${\bf Q}$ is not commensurate as long as the deviation from the commensurate one is not very large.
}

Next, let us discuss the effect of 3D F-QCP in which $C_{q}$ in 
Eq.\ (\ref{rashba7.4}) is proportional to $q^{-1}$: $C_{q}\equiv \sqrt{2}k_{\rm F}{\tilde C}/q$, 
with {$q=|{\bf k}-{\bf k}^{\prime}|\simeq\sqrt{2}k_{\rm F}\sqrt{1-\cos\,\theta}$ and 
${\tilde C}$ being constant}.  
Then, after calculations similar to deriving Eq.\ (\ref{eq:M10}), 
$\Gamma_{\nu}^{\rm R}\ks{\mathbf{k},{i}\delta}$ is expressed as 
\begin{equation}
\Gamma_{\nu}^{\rm R}\ks{\mathbf{k},{i}\delta}
= \alpha_{\rm f}(\eta)\gamma_{\nu}({\bf k}),
\label{eq:M13z}
\end{equation}
where $\alpha_{\rm f}$ is given by 
\begin{equation}
\alpha_{\rm f}(\eta)\equiv 
{
\frac{{\bar \Gamma}_{\nu}^{\rm R}\ks{\mathbf{k},{i}\delta}}{\gamma_{\nu}({\bf k})}+
}
2J\int_{-1}^{1}\frac{{d}(\cos\theta)}{2}
\frac{\sqrt{1-\cos\theta}}{{\tilde C}^{*}\sqrt{a}}\int_{1}^{\infty}{d}y
\frac{1}{\left[y+\frac{\sqrt{a(1-\cos\theta)}}{{\tilde C}^{*}}(y^{\,2}-1)\right]^{2}},
\label{eq:M13}
\end{equation} 
where ${\tilde C}^{*}\equiv v_{\rm F}^{\,2}{\tilde C}/A$.  
As shown in Sects.\ 5 {and 6} of Ref.\ \citen{Suppl}, by calculations similar to those 
deriving Eq.\ (\ref{eq:M11}) from Eq.\ (\ref{eq:M10}), 
the $\Gamma_{\nu}^{\rm R}\ks{\mathbf{k},{i}\delta}$ is, 
up to a logarithmic accuracy in $\eta$, reduced to 
\begin{eqnarray}
& &
\alpha_{\rm f}(\eta)\simeq J
\frac{1}{4k_{\rm F}^{\,2}v_{\rm F}^{\,2}}\left\{\,F\left(\frac{4\sqrt{2}m^{*}A}{{\tilde C}}\right)
+{\frac{\eta}{Ak_{\rm F}^{\,2}}\,\left[G\left(\frac{4\sqrt{2}m^{*}A}{{\tilde C}}\right)\,
\log\frac{e\,\eta}{16Ak_{\rm F}^{\,2}}
+H\left(\frac{2\sqrt{2}m^{*}A}{{\tilde C}}\right)\,
\log\frac{e^{\,2}\,\eta}{16Ak_{\rm F}^{\,2}}\right]}
\right.
\nonumber
\\
& &
\left.
\qquad\qquad\qquad\qquad\qquad\qquad\qquad\qquad\qquad\qquad\qquad\qquad\qquad\qquad
{
+\frac{2\sqrt{Ak_{\rm F}^{2}}}{k_{\rm F}v_{\rm F}}\frac{\pi T}{\sqrt{\eta}}
}
\right\},
\label{eq:M14}
\end{eqnarray} 
where the functions $G(x)$ and $H(x)$ are defined by 
\begin{equation}
G(x)\equiv \frac{x^{\,2}}{8(1+x^{\,2})}\left[1+\frac{1}{x}+\log\frac{x}{2}-F(x)\right],
\label{eq:M15a}
\end{equation}
and 
\begin{equation}
H(x)\equiv\frac{x^{\,2}}{2}\int_{1}^{\infty}{d}y\,
\frac{y^{\,2}-1}{\left[y+x\,(y^{\,2}-1)\right]^{3}}.
\label{eq:M15b}
\end{equation}
%%%%%%%%%%%%%%%%%%%%%%%
Both the functions $G(x)$ and $H(x)$ are positive definite. 

The coefficient $\alpha_{\rm f}(\eta)$ increases sharply as the critical 
point is approached, i.e., $\eta\to 0$, so that it exhibits a sharp {cusp} 
as a function of $\eta$.    
As a result, the band energy $\tilde{\xi}_{\mathbf{k}}^{\pm}$ in 
Eq.\ (\ref{rashba8.3}) is approximated as
{
\begin{align}
\tilde{\xi}_{\mathbf{k}}^{\pm}
&\simeq
\left[\xi_{\mathbf{k}}+\Sigma_{\rm I}^{\rm R}(\mathbf{k},0)
\pm
\left|\alpha+\alpha_{\rm f}(\eta)\right|
\left|\gamma\ks{\mathbf{k}}\right|\right]{\tilde z}^{\pm}_{\mathbf{k}},
\label{rashba11.1bAF}
\end{align}
where ${\tilde z}^{\pm}_{\mathbf{k}}$ is defined by the same expression as Eq.\ (\ref{rashba11.2AF})} 
{
with the same expression as $h$ [Eq.\ (\ref{eq:M9b2})], as discussed 
in Sect.\ 6 of Ref.\ \citen{Suppl}: 
\begin{equation}
h\equiv J\left[
-\,\frac{\sqrt{Ak_{\rm F}^{2}}}{(k_{\rm F}v_{\rm F})^{4}}\,
\frac{\pi T}{\sqrt{\eta}}
+\frac{1}{4(k_{\rm F}v_{\rm F})^{3}}\left(\log\,\frac{4Ak_{\rm F}^{2}}{\eta e}\right)
\right].
\label{eq:M9b23dF}
\end{equation}
}
This means that the coupling constant of the Rashba-SO interaction is enhanced 
due to the critical {F} spin fluctuations, resulting in an increase in the splitting of 
the band (0) into the bands $(\pm)$, which we can easily see if we compare Fig. \ref{rashba9.0}(c) 
with Fig. \ref{rashba9.0}(a), in contrast to the AF critical fluctuations where the band splitting 
is suppressed.  This marked difference arises from {the fact 
that $\gamma_{\nu}({\bf k}-{\bf Q})=\gamma_{\nu}({\bf Q})$ for ${\bf Q}=0$ and 
$\gamma_{\nu}({\bf k}-{\bf Q})=-\gamma_{\nu}({\bf Q})$ for ${\bf Q}=(\pi/a,\pi/a,\pi/c)$, 
because of Eq.\ (\ref{rashba4.3}). 
As in the case of 3D-AFQCP, we expect that the effect of dynamical mass renormalization is 
opposite in two split bands: the effective mass in band (+) is suppressed while that in band 
(-) is enhanced, which is in contrast with the case of 3D AF-QCP. }

We note that the singular behaviors in Eqs.\ (\ref{eq:M11}) and 
(\ref{eq:M14}) become more prominent in two dimensions (2D) than in the 3D case.  
These singularities are obtained by replacing  
the integration with respect to $\cos\theta$ in Eqs.\ (\ref{eq:M10}) and (\ref{eq:M13}) by 
$\int_{0}^{2\pi}d\varphi/2\pi$ and $(1-\cos\theta)$ by $(1-\cos\varphi)$ with $\varphi$ being the 
angle between ${\bf k}^{*}$ and ${\bf k}^{\prime}$.  
As shown in Sects.\ 7 and 9 of Ref.\ \citen{Suppl}, in the case of the AF critical 
point we obtain
\begin{equation}
\alpha_{\rm af}(\eta+Aq_{m}^{\,2})
=\frac{J}{2k_{\rm F}^{\,2}v_{\rm F}^{\,2}}
{\left(\frac{2m^{*}A}{\pi v_{\rm F}C}
\log\frac{64Ak_{\rm F}^{\,2}}{\eta+Aq_{m}^{\,2}}
+\frac{4Ak_{\rm F}^{2}}{k_{\rm F}v_{\rm F}}\frac{T}{\eta+Aq_{m}^{\,2}}
\right)}
\label{eq:M16}
\end{equation}
instead of Eq.\ (\ref{eq:M11}), and {as shown in Sects. 8 and 10 of Ref.\ 23;} 
in the case of the FM critical point, instead of Eq.\ (\ref{eq:M14}), we obtain~\cite{Suppl}
\begin{equation}
\alpha_{\rm f}(\eta)
=\frac{J}{2k_{\rm F}^{\,2}v_{\rm F}^{\,2}}\left[2I\left(\frac{\sqrt{2}m^{*}A}{{\tilde C}}\right)
-\frac{2\sqrt{2A}}{\pi {\tilde C}v_{\rm F}}\sqrt{\eta}\,
{+\frac{2Ak_{\rm F}^{2}}{k_{\rm F}v_{\rm F}}\frac{T}{\eta}}+{\cal O}(\eta)
\right],
\label{eq:M17}
\end{equation}
where the function $I(x)$ is defined by 
\begin{equation}
I(x)\equiv x\,\int_{1}^{\infty}{d}y\,\frac{1}{y^{\,3/2}}\frac{y+x\,(y^{\,2}-1)}
{[y+2x\,(y^{\,2}-1)]^{3/2}}. 
\label{eq:M18}
\end{equation} 
{In the cases of both 2D-AFQCP and 2D-FQCP, the coefficient $h$, 
in the renormalization amplitude ${\tilde z}^{\pm}_{\mathbf{k}}$ [Eq.\ (\ref{rashba11.2AF})], 
is replaced by  
\begin{equation}
h\equiv 
J\left[
-\,\frac{Ak_{\rm F}^{2}}
{(k_{\rm F}v_{\rm F})^{4}}\,\frac{T}{{\eta}}
+\frac{\sqrt{Ak_{\rm F}^{2}}}{4(k_{\rm F}v_{\rm F})^{3}}\frac{1}{\sqrt{\eta}}
\right], 
\label{eq:2dM9b2}
\end{equation}
as shown in Sects.\ 9 and 10 of Ref.\ \citen{Suppl}. }

{
Namely, the extent of singularities in $\alpha_{\rm af}$, $\alpha_{\rm f}$, 
and $h$ is strengthened compared with the 3d case 
because $\eta\propto -T/\log T$ in 2D-AFQCP$\,$~\cite{Moriya2} and $\eta\propto -T\log T$ 
in 2D-FQCP~\cite{Hatatani}.  
In particular, in the case of 2D-AFQCP, the correction to the Rashba-type coupling constant on 
the hot line ($q_{m}=0$) 
diverges logarithmically as the critical point is approached, which 
suggests that the band splitting due to the Rashba-type SO interaction becomes vanishingly 
small on the hot line.  }

In conclusion, on the basis of perturbation in CSFs, we have shown that 
{there exists the off-diagonal (in spin space) self-energy of 
quasiparticles, which gives rise to a considerable renormalization of the 
ASSO interaction.  As a result, the ASSO has been shown to be decreased 
near the hot line(s) on the FS near the AF-QCP while it is uniformly 
increased over the whole FS near the F-QCP, leading to the deformation of the 
FS in different manners for AF-CSF and F-CSF. We have also found} 
that the effective masses of quasiparticles in the two split bands are 
influenced by the ASSO interaction in opposite directions. These predictions 
can be observed through the dHvA experiment in principle. Higher-order effects in 
CSFs, and a self-consistent treatment of the ASSO interaction and CSFs, 
deserve further investigations. {This type of effect on the 
ASSO interaction arising from enhanced spin fluctuations may be possible 
in other physical situations, which are also interesting subjects to 
be explored in future studies.}

%\section{Acknowledgments}
This work is supported by a Grant-in-Aid for Scientific Research 
on Innovative Areas ``Topological Quantum Phenomena'' (No. 22103003) and 
``Ultra Slow Muon Microscope" (No. 23108004) from the Ministry of Education, Culture, 
Sports, Science and Technology, Japan.  
Two of us (K.M. and H.M.) are supported by Grants-in-Aid for Scientific Research 
from the Japan Society for the Promotion of Science (No. 25400369, and Nos. 25220803
and 24244053, respectively).}

\end{document}

\maketitle
%\parindent=1zw

{
\section{Derivation of Eqs.\ (12) and (13) in the text}
Off-diagonal components of fermionic Matsubara Green function matrix in spin space, 
${\hat G}(\mathbf{k},\tau)$, are defined by 
$G_{\uparrow\downarrow}(\mathbf{k},\tau)\equiv -\langle a_{\mathbf{k}\uparrow}(\tau)
a^{\dagger}_{\mathbf{k}\downarrow}(0)\rangle$ and 
$G_{\downarrow\uparrow}(\mathbf{k},\tau)\equiv -\langle a_{\mathbf{k}\downarrow}(\tau)
a^{\dagger}_{\mathbf{k}\uparrow}(0)\rangle$, for $0<\tau<\beta$.   
Then, it is easy to show that the following relation holds:
\begin{equation}
G_{\downarrow\uparrow}(\mathbf{k},\tau)=\left[G_{\uparrow\downarrow}(\mathbf{k},\tau)\right]^{*}.
\label{eq:Green1}
\end{equation}
The Fourier transform of the off-diagonal components, $G_{\downarrow\uparrow}(\mathbf{k},\tau)$ and 
$G_{\uparrow\downarrow}(\mathbf{k},\tau)$, are defined by 
\begin{equation}
G_{\downarrow\uparrow}(\mathbf{k},{\rm i}\varepsilon_{n})\equiv
\int_{0}^{\beta}{\rm d}\tau\ e^{\,{\rm i}\varepsilon_{n}\tau}G_{\downarrow\uparrow}(\mathbf{k},\tau),
\label{eq:Green2}
\end{equation}
and 
\begin{equation}
G_{\uparrow\downarrow}(\mathbf{k},{\rm i}\varepsilon_{n})\equiv
\int_{0}^{\beta}{\rm d}\tau\ e^{\,{\rm i}\varepsilon_{n}\tau}G_{\uparrow\downarrow}(\mathbf{k},\tau).
\label{eq:Green3}
\end{equation} 
With the use of Eq.\ (\ref{eq:Green1}), it is easy to show that the following relation holds:  
\begin{equation}
G_{\downarrow\uparrow}(\mathbf{k},{\rm i}\varepsilon_{n})
=\left[G_{\uparrow\downarrow}(\mathbf{k},-{\rm i}\varepsilon_{n})\right]^{*}.
\label{eq:H2b}
\end{equation}
This is nothing but the relation Eq.\ (13) in the text.  

Since the inverse matrix of ${\hat G}(\mathbf{k},{\rm i}\varepsilon_{n})^{-1}$ satisfies the same 
relation as Eq.\ (\ref{eq:H2b}), the off-diagonal components of the matrix self-energy,
${\hat \Sigma}(\mathbf{k},{\rm i}\varepsilon_{n})$, satisfies the same relation as 
Eq.\ (\ref{eq:H2b}).  Then, considering the relation 
$(\sigma_{\nu})_{\downarrow\uparrow}=(\sigma_{\nu})^{*}_{\uparrow\downarrow}$, Eq.\ (12) in the text 
is immediately derived.  
}

\section{Derivation of Eq.\ (25) in the text}
{
Let us introduce $S_{3{\rm dAF}}(\eta)$ representing double integration part in Eq.\ (24) in the text.    
Substituting the expression (15) in the text into the expression of $S_{3{\rm dAF}}(\eta)$, 
$S_{3{\rm dAF}}(\eta)$
}
is explicitly written as 
\begin{equation}
S_{3{\rm dAF}}(\eta)=
\int_{-1}^{1}\frac{{\rm d}x}{2}
\frac{1}{\sqrt{{\tilde \eta}+K(1-x)}}
\int_{1}^{\infty}{\rm d}y\,
\frac{1}{\left[\frac{1}{C^{*}}\sqrt{{\tilde \eta}+K(1-x)}\,(y^{\,2}-1)+y\right]^{2}},
\label{SM:1}
\end{equation}
where $x=\cos\theta$, ${\tilde \eta}\equiv v_{\rm F}^{\,2}\eta/A$ and 
$K\equiv 2k_{\rm F}^{\,2}v_{\rm F}^{\,2}$. 
Changing the integration variable from $x$ to $u\equiv \sqrt{{\tilde \eta}+K(1-x)}$, Eq.\ (\ref{SM:1}) 
is reduced to 
\begin{equation}
S_{3{\rm dAF}}(\eta)=\frac{1}{K}
\int_{\sqrt{{\tilde \eta}}}^{\sqrt{{\tilde \eta}+2K}}{\rm d}u
\int_{1}^{\infty}{\rm d}y\,
\frac{1}{\left[\frac{u}{C^{*}}(y^{\,2}-1)+y\right]^{2}}.
\label{SM:2}
\end{equation}
Then, the integration with respect to $u$ is easily performed: 
\begin{equation}
S_{3{\rm dAF}}(\eta)=-\frac{C^{*}}{K}
\int_{1}^{\infty}{\rm d}y\,
\frac{1}{y^{\,2}-1}\left[\frac{1}{B_{1}(y^{\,2}-1)+y}-\frac{1}{B_{0}(y^{\,2}-1)+y}\right],
\label{SM:3}
\end{equation}
where $B_{0}$ and $B_{1}$ are defined as follows: 
\begin{eqnarray}
& &B_{0}\equiv \frac{\sqrt{{\tilde \eta}}}{C^{*}}
\\
\label{SM:4a}
& &B_{1}\equiv \frac{\sqrt{{\tilde \eta}+2K}}{C^{*}}
\label{SM:4b}
\end{eqnarray}
Then, the integrand in Eq.\ (\ref{SM:3}) is transformed into  
\begin{equation}
S_{3{\rm dAF}}(\eta)=-\frac{C^{*}}{K}
\int_{1}^{\infty}{\rm d}y\,
\left[\frac{B_{0}}{y}\frac{1}{B_{0}(y^{\,2}-1)+y}-\frac{B_{1}}{y}\frac{1}{B_{1}(y^{\,2}-1)+y}\right]. 
\label{SM:5}
\end{equation}
There holds the following definite integral:
\begin{equation}
\int_{1}^{\infty}{\rm d}y\,
\frac{B}{y}\frac{1}{B\,(y^{\,2}-1)+y}=
\frac{1}{2}\left(\log\,B-\frac{1}{\sqrt{1+4B^{\,2}}}
\log\,\frac{2B+1-\sqrt{1+4B^{\,2}}}{2B+1+\sqrt{1+4B^{\,2}}}\right). 
\label{SM:6}
\end{equation}
With the use of definition (26) in the text, this definite integral is expressed as 
\begin{equation}
\int_{1}^{\infty}{\rm d}y\,
\frac{B}{y}\frac{1}{B\,(y^{\,2}-1)+y}=\frac{1}{2}F(2B). 
\label{SM:7}
\end{equation}
Therefore, Eq.\ (\ref{SM:5}) is expressed as 
\begin{equation}
S_{3{\rm dAF}}(\eta)=-\frac{C^{*}}{2K}\left[ F(2B_{0})-F(2B_{1})\right]. 
\label{SM:8}
\end{equation}
In the limit of $\eta\to 0$, $B_{0}\to 0$ and $B_{1}\to \sqrt{2K}/C^{*}\equiv B^{*}$. 
Then, 
\begin{equation}
\lim_{\eta\to 0}S_{3{\rm dAF}}(\eta)=-\frac{C^{*}}{2K}\left[ F(0)-F(2B^{*})\right]. 
\label{SM:9}
\end{equation}
It is easily shown that $F(0)=0$, and $B^{*}=2m^{*}A/C$ since $K=2k_{\rm F}^{\,2}v_{\rm F}^{\,2}$ and 
$C^{*}=v_{\rm F}^{\,2}C/A$.  Therefore, 
\begin{equation}
\lim_{\eta\to 0}S_{3{\rm dAF}}(\eta)=\frac{C^{*}}{2K}F\left(\frac{4m^{*}A}{C}\right). 
\label{SM:10}
\end{equation}
This gives the first term in the bracket of Eq.\ (25) in the text. The second term 
arises from $F(2B_{0})$ but $F(2B_{1})$ gives only the term of ${\cal O}(\eta)$ 
except for 
{the first term of ${\cal O}(\eta^{\,0})$.
}  
Namely, 
\begin{eqnarray}
& &F(2B_{0})=\log\,B_{0}-\frac{1}{\sqrt{1+4B_{0}^{\,2}}}
\log\,\frac{2B_{0}+1-\sqrt{1+4B_{0}^{\,2}}}{2B_{0}+1+\sqrt{1+4B_{0}^{\,2}}}
\nonumber
\\
& &
\qquad\quad\simeq
{-}\log\,\left(\frac{1-B_{0}}{1+B_{0}}\right)
\simeq 2B_{0}. 
\label{SM:11}
\end{eqnarray}
Then, substituting the definition (\ref{SM:4a}) into Eq.\ (\ref{SM:8}), 
we obtain
\begin{eqnarray}
& &S_{3{\rm dAF}}(\eta)-S_{3{\rm dAF}}(0)={-\frac{C^{*}}{2K}F(2B_{0})
\simeq -\frac{C^{*}}{2K}2B_{0}}
\nonumber
\\
& &\qquad\qquad\qquad\qquad\,\,
{=-\frac{\sqrt{\tilde \eta}}{K}\simeq-\frac{v_{\rm F}}{K\sqrt{A}}\sqrt{\eta}
+{\cal O}(\eta), }
\label{SM:12}
\end{eqnarray}
and the second term in the bracket of Eq.\ (25) in the text.

{
\section{Derivation of Eqs.\ (27) and (28) in the text}
Here we derive Eqs.\ (27) and (28) in the text for 3D-AFQCP.  
Neglecting the $b^{\,2}$ terms in Eq.\ (20) in the text, 
$\Gamma_{\nu}\ks{\mathbf{k},\rm{i}\varepsilon_{n}}$ is expressed as 
\begin{eqnarray}
& &
\Gamma_{\nu}\ks{\mathbf{k},\rm{i}\varepsilon_{n}}
=J\gamma_{\nu}({\bf k}-{\bf Q})\,\int_{-1}^{1}\frac{{\rm d}(\cos\theta)}{2}
\nonumber
\\
& &\qquad\qquad\qquad\qquad
\pi 
T\sum_{n^{\,\prime}\ge 0}\left\{
\frac{1}{\sqrt{{\tilde a}_{+}}}
\frac{1}{(\sqrt{{\tilde a}_{+}}+\varepsilon_{n^{\,\prime}})^{2}}
+
\frac{1}{\sqrt{{\tilde a}_{-}}}
\frac{1}{(\sqrt{{\tilde a}_{-}}+\varepsilon_{n^{\,\prime}})^{2}}
\right.
\nonumber
\\
& &
\left.
\qquad\qquad\qquad\qquad\qquad
+2{\rm i}b
\left[
\frac{1}{\sqrt{{\tilde a}_{+}}}
\frac{1}{(\sqrt{{\tilde a}_{+}}+\varepsilon_{n^{\,\prime}})^{3}}
-
\frac{1}{\sqrt{{\tilde a}_{-}}}
\frac{1}{(\sqrt{{\tilde a}_{-}}+\varepsilon_{n^{\,\prime}})^{3}}
\right]
\right\}, 
\label{eq:G01}
\end{eqnarray} 
where ${\tilde a}_{\pm}\equiv a+C^{*}|\pm\varepsilon_{n}-\varepsilon_{n^{\,\prime}}|$ with 
$a$ given by Eq.\ (15) in the text as 
$a=(v_{\rm F}^{\,2}\eta/A)+2k_{\rm F}^{\,2}v_{\rm F}^{\,2}(1-\cos\,\theta)$, and 
$b\equiv k_{\rm F}v_{\rm F}(1-\cos\,\theta)$, Eq.\ (16) in the text.  
Here, we have neglected terms of ${\cal O}(b^{2})$ 
because the singular contribution to $\Gamma_{\nu}\ks{\mathbf{k},\rm{i}\varepsilon_{n}}$ arises 
from the region $\cos\,\theta\simeq 1$. 
{
We also note that the expression (\ref{eq:G01}) is valid irrespective of the sign of $\varepsilon_{n}$, 
positive or negative. 
}

Since we discuss the retarded function of low-energy quasiparticles, we need a linear dependence of 
$\Gamma_{\nu}\ks{\mathbf{k},\rm{i}\varepsilon_{n}}$ for $\varepsilon_{n}>0$.  
In the limit $T\to 0$, the summation over $\varepsilon_{n^{\,\prime}}$ is approximated by 
integration with respect to $\varepsilon^{\prime}$ as in Eq.\ (21) in the text, except for the term 
$\varepsilon_{n^{\,\prime}}=\varepsilon_{n}$ which corresponds to taking static limit 
($\omega_{m}=0$) of spin-fluctuation 
propagator given by Eq.\ (9) in the text.  The latter term is} {usually} 
{ neglected since it usually gives 
no contribution in the limit $T\to 0$.  
However, it turns out that the most singular contribution 
to the $\varepsilon_{n}$ dependence arises as shown below. 

Let us define the term arising from $\varepsilon_{n^{\,\prime}}=\varepsilon_{n}$ in}
{the first and the third terms in the brace of} 
{Eq.\ (\ref{eq:G01})}
{ (or Eq.\ (18) in the text) } 
{
as ${\bar \Gamma}_{\nu}\ks{\mathbf{k},\rm{i}\varepsilon_{n}}$.  Then, 
}
{
\begin{equation}
{\bar \Gamma}_{\nu}\ks{\mathbf{k},\rm{i}\varepsilon_{n}}
=J\gamma_{\nu}({\bf k}-{\bf Q})\,\pi T\,\int_{-1}^{1}\frac{{\rm d}(\cos\theta)}{2}
\left[
\frac{1}{\sqrt{a}}\frac{1}{(\sqrt{a}+\varepsilon_{n})^{\,2}}
+2{\rm i}b\,
\frac{1}{\sqrt{a}}
\frac{1}{(\sqrt{a}+\varepsilon_{n})^{\,3}}
\right].
\label{eq:G02}
\end{equation} 
The expression in the bracket is expanded in $\varepsilon_{n}$ up to linear order as follows:
\begin{equation}
\left[
\frac{1}{a^{3/2}}-\frac{2}{a^{2}}\varepsilon_{n}
+2{\rm i}\frac{b}{a^{2}}
-2{\rm i}b\frac{3}{a^{5/2}}\varepsilon_{n}+{\cal O}(\varepsilon_{n})^{2}
\right].
\label{eq:G03}
\end{equation} 
}
{
With the use of $a={\tilde \eta}+K(1-x)$ and $b=\sqrt{K/2}(1-x)$, with 
$x\equiv \cos\,\theta$, ${\tilde \eta}\equiv (v_{\rm F}^{\,2}\eta/A)$, and 
$K\equiv 2k_{\rm F}^{\,2}v_{\rm F}^{\,2}$, Eq.\ (\ref{eq:G02}) is reduced to 
}
{
\begin{eqnarray}
& &
{\bar \Gamma}_{\nu}\ks{\mathbf{k},\rm{i}\varepsilon_{n}}
=J\gamma_{\nu}({\bf k}-{\bf Q})\,\pi T\,
\,\int_{-1}^{1}\frac{{\rm d}x}{2}
\left\{\frac{1}{[{\tilde \eta}+K(1-x)]^{3/2}}-
\frac{2\varepsilon_{n}}{[{\tilde \eta}+K(1-x)]^{2}}
\right.
\nonumber
\\
& &
\left.
\qquad\qquad\qquad\qquad\qquad
+2{\rm i}\sqrt{\frac{K}{2}}\frac{(1-x)}{[{\tilde \eta}+K(1-x)]^{2}}
-6{\rm i}\sqrt{\frac{K}{2}}\frac{(1-x)\varepsilon_{n}}{[{\tilde \eta}+K(1-x)]^{5/2}}
\right\}.
\label{eq:G04}
\end{eqnarray}
The integrations with respect to $x$ in Eq.\ (\ref{eq:G04}) are easily performed to give 
\begin{equation}
{\bar \Gamma}_{\nu}\ks{\mathbf{k},\rm{i}\varepsilon_{n}}\simeq
J\gamma_{\nu}({\bf k}-{\bf Q})\,\frac{\pi T}{K}
\left(\frac{1}{\sqrt{{\tilde \eta}}}
-\,\frac{\varepsilon_{n}}{{\tilde \eta}}
+\frac{{\rm i}}{\sqrt{2K}}\,
\log\frac{2K}{{\tilde \eta}}
-\frac{4{\rm i}}{\sqrt{2K}}\,
\frac{\varepsilon_{n}}{\sqrt{{\tilde \eta}}}
\right),
\label{eq:G05}
\end{equation}
where we have retained the most singular term in each contribution in the curly bracket.  
Substituting ${\tilde \eta}\equiv v_{\rm F}^{\,2}\eta/A$ 
and $K\equiv 2k_{\rm F}^{\,2}v_{\rm F}^{\,2}$ into Eq.\ (\ref{eq:G05}), we obtain
\begin{equation}
{\bar \Gamma}_{\nu}\ks{\mathbf{k},\rm{i}\varepsilon_{n}}
\simeq
J\gamma_{\nu}({\bf k}-{\bf Q})\,
\left(
\frac{\sqrt{Ak_{\rm F}^{2}}}{2(k_{\rm F}v_{\rm F})^{3}}\frac{\pi T}{\sqrt{\eta}}
-\,\varepsilon_{n}\,\frac{Ak_{\rm F}^{2}}{2(k_{\rm F}v_{\rm F})^{4}}
\frac{\pi T}{\eta}
+\frac{{\rm i}\,\pi T}{4(k_{\rm F}v_{\rm F})^{3}}\,\log\frac{4Ak_{\rm F}^{2}}{\eta}
-{\rm i}\varepsilon_{n}\,\frac{\sqrt{Ak_{\rm F}^{2}}}{(k_{\rm F}v_{\rm F})^{4}}\,
\frac{\pi T}{\sqrt{\eta}}
\right).
\label{eq:G06}
\end{equation}
Since $\eta\propto T^{3/2}$ just at the 3d-AFQCP,~\cite{MoriyaTakimoto} 
the first and the fourth terms in Eq.\ (\ref{eq:G06}) vanishes as $T^{1/4}$ while the second 
term diverges as $T^{-1/2}$. }
{
Although the first term in Eq.\ (\ref{eq:G06}) vanishes as $T^{1/4}$ in the limit $T\to 0$ at the 
criticality, this $T$ dependence gives a sharper cusp than the second term of Eq.\ (25) in the text 
because $\sqrt{\eta}\propto T^{3/4}$ there.  

The singular behavior as Eq.\ (\ref{eq:G06}), which arises from 
the term $\varepsilon_{n^{\,\prime}}=\varepsilon_{n}$ 
in}
{the first and the third terms in the brace of} 
{
 Eq.\ (\ref{eq:G01}), does not appear for the self-energy 
$\Sigma_{\rm I}({\bf k},{\rm i}\varepsilon_{n})$ 
in Eq.\ (32) in the text.  This is because the denominators 
$[b\pm{\rm i}(\sqrt{{\tilde a}_{\pm}-b^{\,2}}+\varepsilon_{n^{\,\prime}})]^{2}$ in 
Eq.\ (20) in the text should be replaced by 
$[b\pm{\rm i}(\sqrt{{\tilde a}_{\pm}-b^{\,2}}+\varepsilon_{n^{\,\prime}})]$ 
for $\Sigma_{\rm I}({\bf k},{\rm i}\varepsilon_{n})$ so that 
$(\sqrt{a}+\varepsilon_{n})^{\,2}$ and $(\sqrt{a}+\varepsilon_{n})^{\,3}$ are 
replaced by $(\sqrt{a}+\varepsilon_{n})^{\,1}$ and 
$(\sqrt{a}+\varepsilon_{n})^{\,2}$, respectively, in the 
denominator of Eq.\ (\ref{eq:G02}), so that the contribution vanishes in the limit $T\to 0$.  
}

{
The summation in Eq.\ (\ref{eq:G01}) over $\varepsilon_{n^{\,\prime}}$ (except for the term 
$\varepsilon_{n^{\,\prime}}=\varepsilon_{n}$
}
{in the first and the third terms in the brace} 
{) can be approximated, in the region $T\to 0$, by 
integration with respect to $\varepsilon^{\prime}$ as follows:
\begin{equation}
\Gamma_{\nu}\ks{\mathbf{k},\rm{i}\varepsilon_{n}}
-{\bar \Gamma}_{\nu}\ks{\mathbf{k},\rm{i}\varepsilon_{n}}
=J\gamma_{\nu}({\bf k}-{\bf Q})\,\int_{-1}^{1}\frac{{\rm d}(\cos\theta)}{2}
\left[
L(\varepsilon_{n},\cos\theta)+2{\rm i}b\,M(\varepsilon_{n},\cos\theta)
\right],
\label{eq:G1}
\end{equation}
where $L(\varepsilon_{n},\cos\theta)$ and $M(\varepsilon_{n},\cos\theta)$ are given by
}
{
\begin{eqnarray}
& &
L(\varepsilon_{n},\cos\theta)=
\int_{0}^{\varepsilon_{n}-\pi T}\frac{{\rm d}\varepsilon^{\prime}}{2}
\frac{1}{\sqrt{a+C^{*}(\varepsilon_{n}-\varepsilon^{\prime})}
(\sqrt{a+C^{*}(\varepsilon_{n}-\varepsilon^{\prime})}+\varepsilon^{\prime})^{2}}
\nonumber 
\\
& &\quad\qquad\quad\qquad
+\int_{\varepsilon_{n}+\pi T}^{\infty}\frac{{\rm d}\varepsilon^{\prime}}{2}
\frac{1}{\sqrt{a+C^{*}(\varepsilon^{\prime}-\varepsilon_{n})}
(\sqrt{a+C^{*}(\varepsilon^{\prime}-\varepsilon_{n})}+\varepsilon^{\prime})^{2}}
\nonumber 
\\
& &\quad\qquad\quad\qquad
+\int_{0}^{\infty}\frac{{\rm d}\varepsilon^{\prime}}{2}
\frac{1}{\sqrt{a+C^{*}(\varepsilon^{\prime}+\varepsilon_{n})}
(\sqrt{a+C^{*}(\varepsilon^{\prime}+\varepsilon_{n})}+\varepsilon^{\prime})^{2}},
\label{eq:G2a}
\end{eqnarray}
and 
\begin{eqnarray}
& &
M(\varepsilon_{n},\cos\theta)=
\int_{0}^{\varepsilon_{n}-\pi T}\frac{{\rm d}\varepsilon^{\prime}}{2}
\frac{1}{\sqrt{a+C^{*}(\varepsilon_{n}-\varepsilon^{\prime})}
(\sqrt{a+C^{*}(\varepsilon_{n}-\varepsilon^{\prime})}+\varepsilon^{\prime})^{3}}
\nonumber 
\\
& &\quad\qquad\quad\qquad
+\int_{\varepsilon_{n}+\pi T}^{\infty}\frac{{\rm d}\varepsilon^{\prime}}{2}
\frac{1}{\sqrt{a+C^{*}(\varepsilon^{\prime}-\varepsilon_{n})}
(\sqrt{a+C^{*}(\varepsilon^{\prime}-\varepsilon_{n})}+\varepsilon^{\prime})^{3}}
\nonumber 
\\
& &\quad\qquad\quad\qquad
-\int_{0}^{\infty}\frac{{\rm d}\varepsilon^{\prime}}{2}
\frac{1}{\sqrt{a+C^{*}(\varepsilon^{\prime}+\varepsilon_{n})}
(\sqrt{a+C^{*}(\varepsilon^{\prime}+\varepsilon_{n})}+\varepsilon^{\prime})^{3}}.
\label{eq:G2b}
\end{eqnarray}
}

{
In deriving Eqs.\ (\ref{eq:G2a}) and (\ref{eq:G2b}) from Eq.\ (\ref{eq:G01}), 
we have used the relation between a definite integral with respect to $\varepsilon^{\prime}$ and 
a summation over $n^{\,\prime}$ in the region $T\to 0$ such as
}
{
\begin{eqnarray}
& &
\pi T\sum_{0\le n^{\,\prime}<n}F(\varepsilon_{n^{\,\prime}};\varepsilon_{n},T)
+\pi T\sum_{n^{\,\prime}>n}F(\varepsilon_{n^{\,\prime}};\varepsilon_{n},T)
\nonumber
\\
& &
\qquad
\approx\left[\frac{1}{2}\int_{0}^{\varepsilon_{n}-\pi T}{\rm d}\varepsilon^{\prime}
F(\varepsilon^{\prime};\varepsilon_{n},T)
+\frac{1}{2}\int_{\varepsilon_{n}+\pi T}^{\infty}{\rm d}\varepsilon^{\prime}
F(\varepsilon^{\prime};\varepsilon_{n},T)\right]
\left[1+{\cal O}\left(\frac{C^{*}}{a^{3/2}}\,2\pi T\,\varepsilon_{n}\right)\right],
\label{DefiniteIntegral1}
\end{eqnarray}
and 
\begin{equation}
\pi T\sum_{n^{\,\prime}\ge 0}F(\varepsilon_{n^{\,\prime}};\varepsilon_{n},T)
\approx\frac{1}{2}\int_{0}^{\infty}{\rm d}\varepsilon^{\prime}
F(\varepsilon^{\prime};\varepsilon_{n},T)
\left[1+{\cal O}\left(\frac{C^{*}}{a^{3/2}}\,2\pi T\,\varepsilon_{n}\right)\right],
\label{DefiniteIntegral2}
\end{equation}
}
{where $F(\varepsilon_{n^{\,\prime}};\varepsilon_{n},T)$ stands for the functional forms 
in the brace of Eq.\ (\ref{eq:G01}).  The explicit $T$ dependence represents 
that of $\eta$ included in the parameter 
$a=(v_{\rm F}^{\,2}\eta/A)+2k_{\rm F}^{\,2}v_{\rm F}^{\,2}(1-\cos\,\theta)$.  
In deriving the approximate relation in Eqs.\ (\ref{DefiniteIntegral1}) and (\ref{DefiniteIntegral2}), 
we have used the trapezoidal rule and the estimation of its relative accuracy. }

{
By changing the integration variable from $\varepsilon^{\prime}$ to 
$\varepsilon^{\prime\prime}\equiv\varepsilon_{n}-\varepsilon^{\prime}$ 
in the first term in Eq.\ (\ref{eq:G2a}), from $\varepsilon^{\prime}$ to 
$\varepsilon^{\prime\prime}\equiv\varepsilon^{\prime}-\varepsilon_{n}$ in the second 
term in Eq.\ (\ref{eq:G2a}), and from $\varepsilon^{\prime}$ to 
$\varepsilon^{\prime\prime}\equiv\varepsilon_{n}+\varepsilon^{\prime}$ in the third term in 
Eq.\ (\ref{eq:G2a}), $L(\varepsilon_{n},\cos\theta)$ is reduced to 
}
{
\begin{eqnarray}
& &
L(\varepsilon_{n},\cos\theta)=
\int_{\pi T}^{\varepsilon_{n}}\frac{{\rm d}\varepsilon^{\prime\prime}}{2}
\frac{1}{\sqrt{a+C^{*}\varepsilon^{\prime\prime}}
(\sqrt{a+C^{*}\varepsilon^{\prime\prime}}+\varepsilon_{n}-\varepsilon^{\prime\prime})^{2}}
\nonumber 
\\
& &\quad\qquad\quad\qquad
+\int_{\pi T}^{\infty}\frac{{\rm d}\varepsilon^{\prime\prime}}{2}
\frac{1}{\sqrt{a+C^{*}\varepsilon^{\prime\prime}}
(\sqrt{a+C^{*}\varepsilon^{\prime\prime}}+\varepsilon^{\prime\prime}+\varepsilon_{n})^{2}}
\nonumber 
\\
& &\quad\qquad\quad\qquad
+\int_{\pi T}^{\infty}\frac{{\rm d}\varepsilon^{\prime\prime}}{2}
\frac{1}{\sqrt{a+C^{*}\varepsilon^{\prime\prime}}
(\sqrt{a+C^{*}\varepsilon^{\prime\prime}}+\varepsilon^{\prime\prime}-\varepsilon_{n})^{2}}
\nonumber 
\\
& &\quad\qquad\quad\qquad
-\int_{\pi T}^{\varepsilon_{n}}\frac{{\rm d}\varepsilon^{\prime\prime}}{2}
\frac{1}{\sqrt{a+C^{*}\varepsilon^{\prime\prime}}
(\sqrt{a+C^{*}\varepsilon^{\prime\prime}}+\varepsilon^{\prime\prime}-\varepsilon_{n})^{2}}.
\label{eq:G3a}
\end{eqnarray}
}
{
It is easy to see that the sum of the second and the third terms of Eq.\ (\ref{eq:G3a}) 
are even functions in $\varepsilon_{n}$.  The summation of the first and the fourth terms in 
Eq.\ (\ref{eq:G3a}) is calculated in the following way.}

{
Let us introduce $\Delta X(\varepsilon_{n})$ representing the summation of the first and the fourth terms 
of Eq.\ (\ref{eq:G3a}).  Then, it is reduced to 
\begin{eqnarray}
& &
\Delta X(\varepsilon_{n})=
\int_{\pi T}^{\varepsilon_{n}}\frac{{\rm d}\varepsilon^{\prime\prime}}{2}
\frac{1}{\sqrt{a+C^{*}\varepsilon^{\prime\prime}}}
\left[\frac{1}
{(\sqrt{a+C^{*}\varepsilon^{\prime\prime}}+\varepsilon_{n}-\varepsilon^{\prime\prime})^{2}}
-
\frac{1}
{(\sqrt{a+C^{*}\varepsilon^{\prime\prime}}-\varepsilon_{n}+\varepsilon^{\prime\prime})^{2}}
\right]
\label{eq:G3a2}
\\
& &\qquad\quad
=-4
\int_{\pi T}^{\varepsilon_{n}}\frac{{\rm d}\varepsilon^{\prime\prime}}{2}
\frac{\varepsilon_{n}-\varepsilon^{\prime\prime}}
{[a+C^{*}\varepsilon^{\prime\prime}-(\varepsilon_{n}-\varepsilon^{\prime\prime})^{2}]^{2}}.
\label{eq:G3a3}
\end{eqnarray}
By changing the integration variable from $\varepsilon^{\prime\prime}$ to 
$y\equiv \varepsilon^{\prime\prime}/\varepsilon_{n}$, we obtain
\begin{equation}
\Delta X(\varepsilon_{n})=-2\varepsilon_{n}^{2}\int_{\pi T/\varepsilon_{n}}^{1}{\rm d}y
\frac{1-y}{[a+C^{*}\varepsilon_{n}y-\varepsilon_{n}^{2}(1-y)^{2}]^{2}}. 
\label{eq:G3a4}
\end{equation}
Here, we note that $\pi T/\varepsilon_{n}$, the lower limit of integration in Eq.\ (\ref{eq:G3a4}),  
must satisfy $\pi T/\varepsilon_{n}\le 1$ because we are discussing the analytic continuation from 
the upper half plane of ${\rm i}\varepsilon_{n}$.  
Then, up to the order of ${\cal O}(\varepsilon_{n}^{2})$, the denominator of Eq.\ (\ref{eq:G3a4}) 
is approximated by $a^{2}$.  Therefore, $\Delta X(\varepsilon_{n})$ is given by 
\begin{eqnarray}
& &
\Delta X(\varepsilon_{n})=-\,\frac{\varepsilon_{n}^{2}}{a^{2}}
\left[1-\frac{2\pi T}{\varepsilon_{n}}+\frac{(\pi T)^{2}}{\varepsilon_{n}^{2}}\right]
\nonumber
\\
& &
\qquad\quad
=-\frac{\varepsilon_{n}^{2}}{a^{2}}+\pi T\,\frac{2}{a^{2}}\varepsilon_{n}
-\frac{(\pi T)^{2}}{a^{2}}
\nonumber
\\
& &
\qquad\quad
\simeq
-\frac{\varepsilon_{n}^{2}}{a^{2}}+\pi T\,\frac{2}{a^{2}}\varepsilon_{n}.
\label{eq:G3a5}
\end{eqnarray}
where the term $(\pi T)^{2}/a^{2}$ has been discarded because it {gives only a negligible 
contribution to Eq.\ (\ref{eq:G01}) in the asymptotic region $T\sim 0$ as far as 
terms up to ${\cal O}(\varepsilon_{n})$ are concerned.}   
Therefore, up to ${\cal O}(\varepsilon_{n})$, $L(\varepsilon_{n},\cos\,\theta)$ is given by 
\begin{equation}
L(\varepsilon_{n},\cos\,\theta)\simeq 
\int_{\pi T}^{\infty}{\rm d}\varepsilon^{\prime\prime}
\frac{1}{\sqrt{a+C^{*}\varepsilon^{\prime\prime}}
(\sqrt{a+C^{*}\varepsilon^{\prime\prime}}+\varepsilon^{\prime\prime})^{2}}
+\pi T\,\frac{2}{a^{2}}\varepsilon_{n}+{\cal O}(\varepsilon_{n}^{\,2}),  
\label{eq:G3a6}
\end{equation}
where the first term is the contribution from the second and the third terms of 
Eq.\ (\ref{eq:G3a}).}

{
Similarly, the expression Eq.\ (\ref{eq:G2b}) is reduced to  
}
{
\begin{eqnarray}
& &
M(\varepsilon_{n},\cos\theta)=
\int_{\pi T}^{\varepsilon_{n}}\frac{{\rm d}\varepsilon^{\prime\prime}}{2}
\frac{1}{\sqrt{a+C^{*}\varepsilon^{\prime\prime}}
(\sqrt{a+C^{*}\varepsilon^{\prime\prime}}+\varepsilon_{n}-\varepsilon^{\prime\prime})^{3}}
\nonumber 
\\
& &\quad\qquad\quad\qquad
+\int_{\pi T}^{\infty}\frac{{\rm d}\varepsilon^{\prime\prime}}{2}
\frac{1}{\sqrt{a+C^{*}\varepsilon^{\prime\prime}}
(\sqrt{a+C^{*}\varepsilon^{\prime\prime}}+\varepsilon^{\prime\prime}+\varepsilon_{n})^{3}}
\nonumber 
\\
& &\quad\qquad\quad\qquad
-\int_{\pi T}^{\infty}\frac{{\rm d}\varepsilon^{\prime\prime}}{2}
\frac{1}{\sqrt{a+C^{*}\varepsilon^{\prime\prime}}
(\sqrt{a+C^{*}\varepsilon^{\prime\prime}}+\varepsilon^{\prime\prime}-\varepsilon_{n})^{3}}
\nonumber 
\\
& &\quad\qquad\quad\qquad
+\int_{\pi T}^{\varepsilon_{n}}\frac{{\rm d}\varepsilon^{\prime\prime}}{2}
\frac{1}{\sqrt{a+C^{*}\varepsilon^{\prime\prime}}
(\sqrt{a+C^{*}\varepsilon^{\prime\prime}}+\varepsilon^{\prime\prime}-\varepsilon_{n})^{3}}.
\label{eq:G3b}
\end{eqnarray}
The summation of the first and the fourth terms in 
Eq.\ (\ref{eq:G3b}) is calculated in parallel to the derivation of Eq.\ (\ref{eq:G3a5}).  
Let us introduce $\Delta Y(\varepsilon_{n})$ representing the summation of the first and the fourth terms 
of Eq.\ (\ref{eq:G3b}).  Then, it is reduced to 
\begin{eqnarray}
& &
\Delta Y(\varepsilon_{n})=
\int_{\pi T}^{\varepsilon_{n}}\frac{{\rm d}\varepsilon^{\prime\prime}}{2}
\frac{1}{\sqrt{a+C^{*}\varepsilon^{\prime\prime}}}
\left[\frac{1}
{(\sqrt{a+C^{*}\varepsilon^{\prime\prime}}+\varepsilon_{n}-\varepsilon^{\prime\prime})^{3}}
+
\frac{1}
{(\sqrt{a+C^{*}\varepsilon^{\prime\prime}}-\varepsilon_{n}+\varepsilon^{\prime\prime})^{3}}
\right]
\nonumber
\\
& &\qquad\quad
=
\int_{\pi T}^{\varepsilon_{n}}\frac{{\rm d}\varepsilon^{\prime\prime}}{2}
\frac{2(a+C^{*}\varepsilon^{\prime\prime})+6(\varepsilon_{n}-\varepsilon^{\prime\prime})^{2}}
{[a+C^{*}\varepsilon^{\prime\prime}-(\varepsilon_{n}-\varepsilon^{\prime\prime})^{2}]^{3}}.
\label{eq:G3b3}
\end{eqnarray}
By changing the integration variable from $\varepsilon^{\prime\prime}$ to 
$y\equiv \varepsilon^{\prime\prime}/\varepsilon_{n}$, we obtain
\begin{equation}
\Delta Y(\varepsilon_{n})=\varepsilon_{n}\int_{\pi T/\varepsilon_{n}}^{1}{\rm d}y
\frac{a+C^{*}\varepsilon_{n}y+3\varepsilon_{n}^{2}(1-y)^{2}}
{[a+C^{*}\varepsilon_{n}y-\varepsilon_{n}^{2}(1-y)^{2}]^{3}}. 
\label{eq:G3b4}
\end{equation}
Then, up to the order of ${\cal O}(\varepsilon_{n})$, the numerator and the denominator of 
Eq.\ (\ref{eq:G3b4}) are approximated by $a$ and $a^{3}$, respectively.  
Therefore, $\Delta Y(\varepsilon_{n})$ is given by 
\begin{eqnarray}
& &
\Delta Y(\varepsilon_{n})=\frac{\varepsilon_{n}}{a^{2}}
\left(1-\frac{\pi T}{\varepsilon_{n}}\right)
\nonumber
\\
& &
\qquad\quad
=\frac{\varepsilon_{n}}{a^{2}}-\frac{\pi T}{a^{2}}.
\label{eq.G3b5}
\end{eqnarray}
 
Therefore, by expanding the second and the third terms of Eq.\ (\ref{eq:G3b}) 
up to ${\cal O}(\varepsilon_{n})$, $M(\varepsilon_{n},\cos\theta)$ is given by  

\begin{equation}
M(\varepsilon_{n},\cos\theta)\simeq
-\frac{\pi T}{a^{2}}+\frac{\varepsilon_{n}}{a^{2}}-
\int_{\pi T}^{\infty}{\rm d}\varepsilon^{\prime\prime}
\frac{3\varepsilon_{n}}
{\sqrt{a+C^{*}\varepsilon^{\prime\prime}}
(\sqrt{a+C^{*}\varepsilon^{\prime\prime}}+\varepsilon^{\prime\prime})^{4}}+{\cal O}(\varepsilon_{n}^{2}).
\label{eq:G5}
\end{equation}
}
{
By changing the integration variable  from $\varepsilon^{\prime\prime}$ 
to $p\equiv\varepsilon^{\prime\prime}/a$, the third term on the r.h.s. of Eq.\ (\ref{eq:G5}) is reduced to  
}
{
\begin{equation}
-\frac{1}{a^{3/2}}\int_{\pi T/a}^{\infty}{\rm d}p\,
\frac{3\varepsilon_{n}}
{\sqrt{1+C^{*}p}\,(\sqrt{1+C^{*}p}+\sqrt{a}p)^{4}}.
\label{eq:G5a}
\end{equation}
}
{
Singularity arising from this term is much smaller than that from the second term on the r.h.s. 
of Eq.\ (\ref{eq:G5}) 
because the singularity stems from the vanishing behavior in 
$a\equiv (v_{\rm F}^{\,2}\eta/A)+2k_{\rm F}^{\,2}v_{\rm F}^{\,2}(1-\cos\,\theta)$ 
near $\cos\,\theta\simeq 1$ and 
$\eta\simeq 0$.}  
{
Therefore, neglecting the third term on the r.h.s. of Eq.\ (\ref{eq:G5}), we obtain 
}
{
\begin{equation}
M(\varepsilon_{n},\cos\theta)\simeq
-\frac{\pi T}{a^{2}}+
\frac{\varepsilon_{n}}{a^{2}}+{\cal O}(\varepsilon_{n}^{2}). 
\label{eq:G5b}
\end{equation}
}

{
Collecting the results of Eqs.\ (\ref{eq:G1}), (\ref{eq:G3a6}), and (\ref{eq:G5b}), we obtain the 
$\Gamma_{\nu}\ks{\mathbf{k},\rm{i}\varepsilon_{n}}$ up to 
${\cal O}(\varepsilon_{n})$ as follows:}
{
\begin{eqnarray}
& &
\Gamma_{\nu}\ks{\mathbf{k},\rm{i}\varepsilon_{n}}
\simeq 
{\bar \Gamma}_{\nu}\ks{\mathbf{k},\rm{i}\varepsilon_{n}}
+J\gamma_{\nu}({\bf k}-{\bf Q})\,\int_{-1}^{1}\frac{{\rm d}(\cos\theta)}{2}
\left[
\int_{\pi T}^{\infty}{\rm d}\varepsilon^{\prime\prime}
\frac{1}{\sqrt{a+C^{*}\varepsilon^{\prime\prime}}
(\sqrt{a+C^{*}\varepsilon^{\prime\prime}}+\varepsilon^{\prime\prime})^{2}}
\right.
\nonumber
\\
& &
\left.
\qquad\qquad\qquad\qquad\qquad\qquad\qquad\qquad\qquad\qquad
+\pi T\,\frac{2}{a^{2}}\varepsilon_{n}+
2{\rm i}b\left(-\frac{\pi T}{a^{2}}+\frac{\varepsilon_{n}}{a^{2}}\right)\right]
+{\cal O}(\varepsilon_{n}^{2}),
\label{eq:G6}
\end{eqnarray}
where $a\equiv(v_{\rm F}^{\,2}\eta/A)+2k_{\rm F}^{\,2}v_{\rm F}^{\,2}(1-\cos\,\theta)$ and 
$b\equiv k_{\rm F}v_{\rm F}(1-\cos\,\theta)$.  
Then, the linear terms of $\Gamma_{\nu}\ks{\mathbf{k},\rm{i}\varepsilon_{n}}$ in $\varepsilon_{n}$ 
is obtained as follows:
\begin{eqnarray}
& &
\Gamma_{\nu}\ks{\mathbf{k},\rm{i}\varepsilon_{n}}
-\Gamma_{\nu}\ks{\mathbf{k},\rm{i}\varepsilon_{n}}|_{\varepsilon_{n}=0}
\simeq 
{\bar \Gamma}_{\nu}\ks{\mathbf{k},\rm{i}\varepsilon_{n}}
-{\bar \Gamma}_{\nu}\ks{\mathbf{k},\rm{i}\varepsilon_{n}}|_{\varepsilon_{n}=0}
\nonumber
\\
& &
\qquad\qquad\qquad\qquad\qquad\qquad\qquad
+J\gamma_{\nu}({\bf k}-{\bf Q})\,\int_{-1}^{1}\frac{{\rm d}(\cos\theta)}{2}
\left[
\pi T\,\frac{2}{a^{2}}\varepsilon_{n}+2\frac{b}{a^{2}}({\rm i}\varepsilon_{n})
\right].
\label{eq:G6b}
\end{eqnarray}
We note that, in the limit $T\to 0$, the first term in the square bracket of Eq.\ (\ref{eq:G6}) is 
reduced to the second term of Eq.\ (22) in the text.  
}
{
Integrations with respect to $\cos\,\theta$ on the r.h.s. of Eq.\ (\ref{eq:G6}) are easily 
performed: }
{
\begin{equation}
\int_{-1}^{1}\frac{{\rm d}(\cos\theta)}{2}
\frac{1}{a^{2}}=
\frac{Ak_{\rm F}^{2}}{4(k_{\rm F}v_{\rm F})^{4}}
\left(\frac{1}{\eta}-\frac{1}{\eta+4Ak_{\rm F}^{2}}\right)
\simeq
\frac{Ak_{\rm F}^{2}}{4(k_{\rm F}v_{\rm F})^{4}}\frac{1}{\eta},
\label{eq:G7a}
\end{equation}
and }
{
\begin{equation}
\int_{-1}^{1}\frac{{\rm d}(\cos\theta)}{2}
\frac{b}{a^{2}}=
\frac{1}{8(k_{\rm F}v_{\rm F})^{3}}
\left(
\log\,\frac{4Ak_{\rm F}^{2}+\eta}{\eta}-\frac{4Ak_{\rm F}^{2}}{4Ak_{\rm F}^{2}+\eta}
\right)
\simeq
\frac{1}{8(k_{\rm F}v_{\rm F})^{3}}\,\log\,\frac{4Ak_{\rm F}^{2}}{\eta e},
\label{eq:G7b}
\end{equation}
}
{
where we have retained the most singular terms and within the logarithmic accuracy  
to obtain the last approximate expressions.  

Finally, substituting Eqs.\ (\ref{eq:G06}), (\ref{eq:G7a}), and (\ref{eq:G7b}) into Eq.\ (\ref{eq:G6b}), 
we obtain
\begin{equation}
\Gamma_{\nu}\ks{\mathbf{k},\rm{i}\varepsilon_{n}}
\simeq
\Gamma_{\nu}\ks{\mathbf{k},\rm{i}\varepsilon_{n}}|_{\varepsilon_{n}=0}
+J\gamma_{\nu}({\bf k}-{\bf Q})\left[
-\frac{\sqrt{Ak_{\rm F}^{2}}}{(k_{\rm F}v_{\rm F})^{4}}
\frac{\pi T}{\sqrt{\eta}}
+\frac{1}{4(k_{\rm F}v_{\rm F})^{3}}\left(\log\,\frac{4Ak_{\rm F}^{2}}{\eta e}\right)
\right]({\rm i}\varepsilon_{n})
+{\cal O}(\varepsilon_{n}^{2}).
\label{eq:G8}
\end{equation}
}
{  
Here we note that the term obtained from the first term in the square bracket of 
Eq.\ (\ref{eq:G6b}) cancels out  
the second term in the bracket of Eq.\ (\ref{eq:G06}), 
so that the term proportional to $\varepsilon_{n}$ disappears but only that proportional to 
${\rm i}\varepsilon_{n}$ remains. 
This result verifies the fact that the first line in the brace of 
Eq.\ (\ref{eq:G01}), i.e., its real part, is an even function in $\varepsilon_{n}$. 
}
{Equation (\ref{eq:G8}) gives Eqs.\ (27) and (28) in the text.} 
{
We also note that the term $-2{\rm i}b\,\pi T/a^{2}$ 
in Eq.\ (\ref{eq:G6}) cancels out the third term in Eq.\ (\ref{eq:G03}) or Eq.\ (\ref{eq:G06}).
}

%C^{*}\left(-\,\varepsilon_{n}\,\frac{(Ak_{\rm F}^{2})^{3/2}}{2(k_{\rm F}v_{\rm F})^{5}}
%\frac{\pi T}{\eta^{3/2}}
%+{\rm i}\varepsilon_{n}\,\frac{Ak_{\rm F}^{2}}{2(k_{\rm F}v_{\rm F})^{5}}\,
%\frac{\pi T}{\eta}
%\right)

%We note that the coefficient $C^{*}$ does not appear in the final expression (\ref{eq:G8}).  This 
%implies that the expression (\ref{eq:G8}) for 3d-AFQCP is valid also in the 3d-FQCP because the 
%difference between AF and F spin fluctuations appears only through the $q$ dependence of $C_{q}$ of 
%the spin-fluctuation propagator, Eq.\ (9) in the text.  Of course, temperature dependence of 
%$\eta$ in 3d-FQCP is different from that in 3d-AFQCP:~\cite{MoriyaTakimoto}  
%$\eta\propto T^{4/3}$ so that the first term in Eq.\ (\ref{eq:G06}) diverges 
%as $T^{-1/3}$ and the second term vanishes as $T^{1/3}$, making the expression (\ref{eq:G8}) valid.  

{
\section{Derivation of Eqs.\ (33) and (34) in the text}
With the use of Eqs.\ (10) and (12) in the text, 
an explicit form of Eq.\ (31) in the text is given by
\begin{equation}
\left[{\hat G}^{\rm R}(\mathbf{k},\varepsilon+{\rm i}\delta)\right]^{-1}
=\left(
\begin{array}{cc}
\varepsilon-\xi_{\mathbf{k}}-\Sigma_{\rm I}^{\rm R}(\mathbf{k},\varepsilon) & 
-\alpha{\bm{\gamma}}\ks{\mathbf{k}}\cdot\mathbf{\hat{\bm{\sigma}}}_{\uparrow\downarrow}
-\Sigma_{\rm{I\hspace{-.1em}I}\uparrow\downarrow}^{\rm R}
\ks{\mathbf{k},\varepsilon}\\
-\alpha{\bm{\gamma}}\ks{\mathbf{k}}\cdot\mathbf{\hat{\bm{\sigma}}}_{\downarrow\uparrow}-
\Sigma_{\rm{I\hspace{-.1em}I}\downarrow\uparrow}^{\rm R}
\ks{\mathbf{k},\varepsilon}&
\varepsilon-\xi_{\mathbf{k}}-\Sigma_{\rm I}^{\rm R}(\mathbf{k},\varepsilon)
\end{array}
\right),
\label{eq:H1}
\end{equation}
where $\Sigma_{\rm{I\hspace{-.1em}I}\uparrow\downarrow}^{\rm R}$ and 
$\Sigma_{\rm{I\hspace{-.1em}I}\downarrow\uparrow}^{\rm R}$ are defined by Eqs.\ (29) and 
(30) in the text, respectively.  For concise presentation, let us introduce the following 
notation: 
\begin{equation}
{\tilde \Gamma}_{\nu}({\bf k},\varepsilon)
\equiv \alpha\gamma_{\nu}(\mathbf{k})+
{\gamma}_{\nu}\ks{\mathbf{k}-\mathbf{Q}}
[\alpha_{\rm {af}}\ks{\eta}+(h+{\rm i}g)\varepsilon].
\label{eq:H2}
\end{equation}
}
{
Although $g=0$ as discussed in Sect. 3 (e.g., as derived from Eq.\ (\ref{eq:G8})), we here give 
a general expression for the dispersion of quasiparticles assuming $g\not=0$ in general.}
{ 
Then, Eq.\ (\ref{eq:H1}) is written in a concise form as follows:
\begin{equation}
\left[{\hat G}^{\rm R}(\mathbf{k},\varepsilon+{\rm i}\delta)\right]^{-1}
=\left(
\begin{array}{cc}
\varepsilon-\xi_{\mathbf{k}}-\Sigma_{\rm I}^{\rm R}(\mathbf{k},\varepsilon) & 
-\sum_{\nu=x,y}{\tilde \Gamma}_{\nu}({\bf k},\varepsilon)\sigma_{\nu}^{\uparrow\downarrow}\\
-\sum_{\nu=x,y}{\tilde \Gamma}^{*}_{\nu}({\bf k},\varepsilon)(\sigma_{\nu}^{\uparrow\downarrow})^{*}&
\varepsilon-\xi_{\mathbf{k}}-\Sigma_{\rm I}^{\rm R}(\mathbf{k},\varepsilon)
\end{array}
\right).
\label{eq:H1a}
\end{equation}

Dispersion of quasiparticles is determined by the condition that 
$|{\hat G}^{\rm R}(\mathbf{k},\varepsilon+{\rm i}\delta)|^{-1}=0 $: 
\begin{equation}
\left[\varepsilon-\xi_{\mathbf{k}}-\Sigma_{\rm I}^{\rm R}(\mathbf{k},\varepsilon)\right]^{2}
=\sum_{\nu=x,y}{\tilde \Gamma}_{\nu}({\bf k},\varepsilon)\sigma_{\nu}^{\uparrow\downarrow}\times
\sum_{\nu=x,y}[{\tilde \Gamma}({\bf k},\varepsilon)]^{*}_{\nu}\sigma_{\nu}^{\downarrow\uparrow}
\label{eq:H3}
\end{equation}
The r.h.s. of Eq.\ (\ref{eq:H3}) is transformed into 
\begin{equation}
\sum_{\nu=x,y}{\tilde \Gamma}_{\nu}\sigma_{\nu}^{\uparrow\downarrow}\times
\sum_{\nu=x,y}{\tilde \Gamma}^{*}_{\nu}\sigma_{\nu}^{\downarrow\uparrow}
=\left(|{\tilde \Gamma}_{x}|^{2}+|{\tilde \Gamma}_{y}|^{2}\right)
+{\rm i}\left({\bm{\tilde \Gamma}}\times{\bm{\tilde \Gamma}}^{*}\right)_{z},
\label{eq:H4}
\end{equation}
where we have abbreviated 
${\tilde \Gamma}_{\nu}({\bf k},\varepsilon)$ to ${\tilde \Gamma}_{\nu}$ for conciseness.  
With the use of Eq.\ (\ref{eq:H2}), we calculate the two terms on the r.h.s. of Eq.\ (\ref{eq:H4}): 
\begin{equation}
|{\tilde \Gamma}_{x}|^{2}+|{\tilde \Gamma}_{y}|^{2}
=\left[\alpha{\bm{\gamma}}\ks{\mathbf{k}}+\alpha_{\rm af}{\bm{\gamma}}\ks{\mathbf{k}-\mathbf{Q}}\right]^{2}
+2h\varepsilon\,{\bm{\gamma}}\ks{\mathbf{k}-\mathbf{Q}}\cdot
\left[\alpha{\bm{\gamma}}\ks{\mathbf{k}}+\alpha_{\rm af}{\bm{\gamma}}\ks{\mathbf{k}-\mathbf{Q}}\right]
+{\cal O}(\varepsilon^{2}),
\label{eq:H5}
\end{equation}
and 
\begin{equation}
{\rm i}\left({\bm{\tilde \Gamma}}\times{\bm{\tilde \Gamma}}^{*}\right)_{z}
=2\alpha g\varepsilon\left[{\bm{\gamma}}\ks{\mathbf{k}}
\times{\bm{\gamma}}\ks{\mathbf{k}-\mathbf{Q}}\right]_{z}.
\label{eq:H6}
\end{equation}
Then, the condition Eq.\ (\ref{eq:H3}), determining the dispersion of quasiparticles, is 
reduced to 
\begin{eqnarray}
& &
\varepsilon-\xi_{\mathbf{k}}-\Sigma_{\rm I}^{\rm R}(\mathbf{k},0)
-\left.\frac{\partial \Sigma_{\rm{I}}^{\rm R}\ks{\mathbf{k},\varepsilon}}
{\partial \varepsilon}\right|_{\varepsilon=0}\varepsilon
\nonumber
\\
& &\qquad\quad
=\pm\left\{
\left[\alpha{\bm{\gamma}}\ks{\mathbf{k}}+\alpha_{\rm af}{\bm{\gamma}}\ks{\mathbf{k}-\mathbf{Q}}\right]^{2}
+2\alpha g\varepsilon\left[{\bm{\gamma}}\ks{\mathbf{k}}
\times{\bm{\gamma}}\ks{\mathbf{k}-\mathbf{Q}}\right]_{z}
\right.
\nonumber
\\
& &
\left.
\qquad\quad\qquad\qquad
+2h\varepsilon\,{\bm{\gamma}}\ks{\mathbf{k}-\mathbf{Q}}\cdot
\left[\alpha{\bm{\gamma}}\ks{\mathbf{k}}+\alpha_{\rm af}{\bm{\gamma}}\ks{\mathbf{k}-\mathbf{Q}}\right]
\right\}^{1/2}.
\label{eq:H7}
\end{eqnarray}

Therefore, the dispersion of the quasiparticles is given by 
\begin{eqnarray}
& &
\varepsilon\left\{1-\left.\frac{\partial \Sigma_{\rm{I}}^{\rm R}\ks{\mathbf{k},\varepsilon}}
{\partial \varepsilon}\right|_{\varepsilon=0}
\mp\frac{h\,[\alpha\bm{\gamma}({\bf k})+\alpha_{\rm af}\bm{\gamma}({\bf k}-{\bf Q})]
\cdot\bm{\gamma}({\bf k}-{\bf Q})}{|\alpha\bm{\gamma}({\bf k})
+\alpha_{\rm af}\bm{\gamma}({\bf k}-{\bf Q})|}
\right\}
\nonumber
\\
& &\qquad\qquad
=
\xi_{\mathbf{k}}+\Sigma_{\rm I}^{\rm R}(\mathbf{k},0)
\pm\left|
\alpha{\bm{\gamma}}\ks{\mathbf{k}}+\alpha_{\rm af}{\bm{\gamma}}\ks{\mathbf{k}-\mathbf{Q}}\right|
\pm 
\frac{\alpha (g\varepsilon)\,
\left[{\bm{\gamma}}\ks{\mathbf{k}}\times{\bm{\gamma}}\ks{\mathbf{k}-\mathbf{Q}}\right]_{z}}
{\left|\alpha{\bm{\gamma}}\ks{\mathbf{k}}+\alpha_{\rm af}{\bm{\gamma}}\ks{\mathbf{k}-\mathbf{Q}}\right|}.
\label{eq:H8}
\end{eqnarray}
This gives the dispersion relation presented as Eqs.\ (33) and (34) in the text} 
{if $g=0$.  
}

\section{Derivation of Eq.\ (38) in the text}
{
Let us introduce $S_{3{\rm dF}}(\eta)$ representing double integration part in Eq.\ (38) in the text.   
Substituting the expression (15) in the text into the expression of $S_{3{\rm dF}}(\eta)$, 
$S_{3{\rm dF}}(\eta)$
}
is explicitly written as 
\begin{equation}
S_{3{\rm dF}}(\eta)=
\int_{-1}^{1}\frac{{\rm d}x}{2}
\frac{\sqrt{1-x}}{{\tilde C}^{*}\sqrt{{\tilde \eta}+K(1-x)}}
\int_{1}^{\infty}{\rm d}y\,
\frac{1}{\left[\frac{\sqrt{1-x}}{{\tilde C}^{*}}\sqrt{{\tilde \eta}+K(1-x)}\,(y^{\,2}-1)+y\right]^{2}},
\label{SM2:1}
\end{equation}
where $x=\cos\theta$, ${\tilde \eta}\equiv v_{\rm F}^{\,2}\eta/A$ and 
$K\equiv 2k_{\rm F}^{\,2}v_{\rm F}^{\,2}$. 
Changing the integration variable from $x$ to $s\equiv \sqrt{1-x}$, Eq.\ (\ref{SM2:1}) 
is reduced to 
\begin{equation}
S_{3{\rm dF}}(\eta)=
\int_{0}^{\sqrt{2}}{\rm d}s\,
\frac{s^{\,2}}{{\tilde C}^{*}\sqrt{{\tilde \eta}+Ks^{\,2}}}
\int_{1}^{\infty}{\rm d}y\,
\frac{1}{\left[\frac{s\sqrt{{\tilde \eta}+Ks^{\,2}}}{{\tilde C}^{*}}(y^{\,2}-1)+y\right]^{2}}.
\label{SM2:2}
\end{equation}
First, we calculate $S_{3{\rm dF}}(0)$ which is given by 
\begin{equation}
S_{3{\rm dF}}(0)=
\int_{0}^{\sqrt{2}}{\rm d}s\,
\frac{s}{{\tilde C}^{*}\sqrt{K}}
\int_{1}^{\infty}{\rm d}y\,
\frac{1}{\left[\frac{\sqrt{K}s^{\,2}}{{\tilde C}^{*}}(y^{\,2}-1)+y\right]^{2}}.
\label{SM2:3}
\end{equation}
Changing the integration variable from $s$ to $z\equiv s^{\,2}$, Eq.\ (\ref{SM2:3}) is reduced to 
\begin{equation}
S_{3{\rm dF}}(0)=\frac{1}{2{\tilde C}^{*}\sqrt{K}}
\int_{0}^{2}{\rm d}z
\int_{1}^{\infty}{\rm d}y\,
\frac{1}{\left[\frac{\sqrt{K}}{{\tilde C}^{*}}z\,(y^{\,2}-1)+y\right]^{2}}.
\label{SM2:4}
\end{equation}
Then the integration with respect to $z$ is easily performed: 
\begin{eqnarray}
& &
S_{3{\rm dF}}(0)=-\frac{1}{2K}
\int_{1}^{\infty}{\rm d}y\,
\frac{1}{y^{\,2}-1}
\left[\frac{1}{\frac{2\sqrt{K}}{{\tilde C}^{*}}(y^{\,2}-1)+y}-\frac{1}{y}\right]
\nonumber
\\
& &\qquad\qquad
=\frac{1}{2K}
\int_{1}^{\infty}{\rm d}y\,
\frac{\frac{2\sqrt{K}}{{\tilde C}^{*}}}
{y\left[\frac{2\sqrt{K}}{{\tilde C}^{*}}(y^{\,2}-1)+y\right]}.
\label{SM2:5}
\end{eqnarray}
Then, with the use of {Eq.\ (\ref{SM:7})}, we obtain
\begin{equation}
S_{3{\rm dF}}(0)=\frac{1}{4K}F\left(\frac{{4}\sqrt{K}}{{\tilde C}^{*}}\right)
=\frac{1}{4K}F\left(\frac{4\sqrt{2}m^{*}A}{{\tilde C}}\right),
\label{SM2:6}
\end{equation}
where we have used definitions, ${\tilde C}^{*}\equiv {\tilde C}v_{\rm F}^{\,2}/A$ and 
$K\equiv 2k_{\rm F}^{\,2}v_{\rm F}^{\,2}$. 

Next, let us discuss the lowest-order correction in $\eta$, which will turn out to be of the order of 
${\cal O}(\eta\log\eta)$.  To this end, we calculate $S_{3{\rm dF}}(\eta)-S_{3{\rm dF}}(0)$: 
With the use of the expression Eq.\ (\ref{SM2:2}), we obtain 
\begin{eqnarray}
& &S_{3{\rm dF}}(\eta)-S_{3{\rm dF}}(0)=\frac{1}{{\tilde C}^{*}}\left\{
\int_{0}^{\sqrt{2}}{\rm d}s\,
\frac{s^{\,2}}{\sqrt{{\tilde \eta}+Ks^{\,2}}}
\int_{1}^{\infty}{\rm d}y\,
\frac{1}{\left[\frac{s\sqrt{{\tilde \eta}+Ks^{\,2}}}{{\tilde C}^{*}}(y^{\,2}-1)+y\right]^{2}}
\right.
\nonumber
\\
& &
\qquad\qquad\qquad\qquad\qquad\qquad
\left.
-\int_{0}^{\sqrt{2}}{\rm d}s\,
\frac{s}{\sqrt{K}}
\int_{1}^{\infty}{\rm d}y\,
\frac{1}{\left[\frac{\sqrt{K}s^{\,2}}{{\tilde C}^{*}}(y^{\,2}-1)+y\right]^{2}}
\right\}
\nonumber
\\
& &
\qquad\quad
=\int_{0}^{\sqrt{2}}{\rm d}s\,\frac{1}{{\tilde C}^{*}}
\left[\frac{s^{\,2}}{\sqrt{{\tilde \eta}+Ks^{\,2}}}-\frac{s}{\sqrt{K}}\right]
\int_{1}^{\infty}{\rm d}y\,
\frac{1}{\left[\frac{s\sqrt{{\tilde \eta}+Ks^{\,2}}}{{\tilde C}^{*}}(y^{\,2}-1)+y\right]^{2}}
\nonumber
\\
& &
\qquad\qquad
+\int_{0}^{\sqrt{2}}{\rm d}s\,
\frac{s}{{\tilde C}^{*}\sqrt{K}}
\int_{1}^{\infty}{\rm d}y\left\{
\frac{1}{\left[\frac{s\sqrt{{\tilde \eta}+Ks^{\,2}}}{{\tilde C}^{*}}(y^{\,2}-1)+y\right]^{2}}
-\frac{1}{\left[\frac{\sqrt{K}s^{\,2}}{{\tilde C}^{*}}(y^{\,2}-1)+y\right]^{2}}
\right\}.
\label{SM2:7}
\end{eqnarray}
It is easy to see that the second term of Eq.\ (\ref{SM2:7}) can be 
expanded into the Taylor series so that this term gives only ${\cal O}(\eta)$ at most and 
can be safely neglected up to the logarithmic accuracy.  Let us define the integration with respect $y$ 
in the first term of Eq.\ (\ref{SM2:7}) as $L(s)$: i.e., 
\begin{equation}
L(s)\equiv
\int_{1}^{\infty}{\rm d}y\,
\frac{1}{\left[\frac{s\sqrt{{\tilde \eta}+Ks^{\,2}}}{{\tilde C}^{*}}(y^{\,2}-1)+y\right]^{2}}.
\label{SM2:8}
\end{equation}
Then, the first term of Eq.\ (\ref{SM2:7}), denoted by $X(\eta)$, is expressed as 
\begin{equation}
X(\eta)=
\frac{1}{{\tilde C}^{*}}\int_{0}^{\sqrt{2}}{\rm d}s
\left[\frac{s^{\,2}}{\sqrt{{\tilde \eta}+Ks^{\,2}}}-\frac{s}{\sqrt{K}}\right]L(s).
\label{SM2:9}
\end{equation}
Changing the integration variable from $s$ to $z\equiv s^{\,2}$, Eq.\ (\ref{SM2:9}) is transformed 
through partial integration with respect to $z$ as follows:
\begin{eqnarray}
& &
X(\eta)=\frac{1}{2{\tilde C}^{*}\sqrt{K}}\int_{0}^{{2}}{\rm d}z
\left(\sqrt{\frac{z}{z+\frac{{\tilde \eta}}{K}}}-1\right)L(\sqrt{z})
\nonumber
\\
& &
\qquad
=\frac{1}{2{\tilde C}^{*}\sqrt{K}}\left\{
\left[\sqrt{2\left(2+\frac{{\tilde \eta}}{K}\right)}-2+
\frac{1}{2}\frac{{\tilde \eta}}{K}\log
\left|
\frac{\sqrt{2+\frac{{\tilde \eta}}{K}}-\sqrt{2}}{\sqrt{2+\frac{{\tilde \eta}}{K}}+\sqrt{2}}
\right|
\right]L(\sqrt{2})
\right.
\nonumber
\\
& &
\qquad\qquad
\left.
-\int_{0}^{2}{\rm d}z
\left[\sqrt{z\left(z+\frac{{\tilde \eta}}{K}\right)}-z+
\frac{1}{2}\frac{{\tilde \eta}}{K}\log\left|
\frac{\sqrt{z+\frac{{\tilde \eta}}{K}}-\sqrt{z}}{\sqrt{z+\frac{{\tilde \eta}}{K}}+\sqrt{z}}\right|
\right]\frac{{\rm d}L(\sqrt{z})}{{\rm d}z}
\right\}.
\label{SM2:10}
\end{eqnarray} 
In the limit of $\eta\to 0$, the first term in the brace of Eq.\ (\ref{SM2:10}), denoted by 
${2\sqrt{K}}X^{(1)}(\eta)$, is given by 
\begin{equation}
X^{(1)}(\eta)\simeq\frac{1}{4}\frac{\tilde \eta}{K^{\,3/2}}\log\frac{e\,{\tilde \eta}}{8K}\times
\lim_{\eta\to 0}L(\sqrt{2}),
\label{SM2:11}
\end{equation} 
where, according to Eq.\ (\ref{SM2:8}), $\lim_{\eta\to 0}L(\sqrt{2})$ is given by 
\begin{equation}
\lim_{\eta\to 0}L(\sqrt{2})=\int_{1}^{\infty}{\rm d}y\,
\frac{1}{\left[\frac{2\sqrt{K}}{{\tilde C}^{*}}(y^{\,2}-1)+y\right]^{2}}.  
\label{SM2:12}
\end{equation} 
After elementary integrations, an explicit form of Eq.\ (\ref{SM2:12}) is 
\begin{equation}
\lim_{\eta\to 0}L(\sqrt{2})=
\frac{2{\tilde B}^{\,*}}{1+4{\tilde B}^{\,*2}}
\left[1+\frac{1}{2{\tilde B}^{\,*}}+\log{\tilde B}^{\,*}-F(2{\tilde B}^{\,*})
\right]= \frac{4}{{\tilde B}^{\,*}}G(2{\tilde B}^{\,*}),
\label{SM2:12a}
\end{equation} 
where 
\begin{equation}
{\tilde B}^{\,*}\equiv \frac{2\sqrt{K}}{{\tilde C}^{*}}. 
\label{SM2:13}
\end{equation}
{
The last equality in Eq.\ (\ref{SM2:12a}) is obtained using the function $G(x)$ 
defined by Eq.\ (40) in the text.
}

The second term in the brace of Eq.\ (\ref{SM2:10}), denoted by 
${2\sqrt{K}}X^{(2)}(\eta)$, is estimated as follows:  
Changing the integration variable from $z$ to $w\equiv Kz/{\tilde \eta}$, $X^{(2)}(\eta)$ is 
expressed as 
\begin{equation}
X^{(2)}(\eta)=
-\frac{{\tilde \eta}}{2K^{\,3/2}}\int_{0}^{2K/{\tilde \eta}}{\rm d}w
\left[\sqrt{w(w+1)}-w+
\frac{1}{2}\log\left|
\frac{\sqrt{w+1}-\sqrt{w}}{\sqrt{w+1}+\sqrt{w}}\right|
\right]\frac{{\rm d}L(\sqrt{{\tilde \eta}w/K})}{{\rm d}w}.
\label{SM2:14}
\end{equation}
Since the upper limit of the integration with respect to $w$ in Eq.\ (\ref{SM2:14}) diverges in the 
limit $\eta \to 0$, there is a possibility that a singularity in $\eta$ arises.  
Therefore, the behavior of ${\rm d}L(\sqrt{{\tilde \eta}w/K})/{\rm d}w$ at $w=2K/{\tilde \eta}\to \infty$ 
should be estimated.  With the use of expression for $L(s)$, Eq.\ (\ref{SM2:8}), we obtain 
\begin{equation}
\frac{{\rm d}L(\sqrt{{\tilde \eta}w/K})}{{\rm d}w}=-
\frac{{\tilde \eta}}{{\tilde C}^{*}\sqrt{K}}
\int_{1}^{\infty}{\rm d}y\,
\frac{y^{\,2}-1}{\left[\frac{{\tilde \eta}\sqrt{w(w+1)}}{{\tilde C}^{*}\sqrt{K}}(y^{\,2}-1)+y\right]^{3}}
\left(\sqrt{\frac{w+1}{w}}+\sqrt{\frac{w}{w+1}}\right).
\label{SM2:15}
\end{equation}
Near the upper limit of $w=2K/{\tilde \eta}$ in the limit of $\eta\to 0$, 
${\rm d}L(\sqrt{{\tilde \eta}w/K})/{\rm d}w$ is estimated as 
\begin{equation}
\frac{{\rm d}L(\sqrt{{\tilde \eta}w/K})}{{\rm d}w}\simeq
-\frac{2{\tilde \eta}}{{\tilde C}^{*}\sqrt{K}}
\int_{1}^{\infty}{\rm d}y\,
\frac{y^{\,2}-1}{\left[\frac{2\sqrt{K}}{{\tilde C}^{*}}(y^{\,2}-1)+y\right]^{3}}.
\label{SM2:16}
\end{equation}
With the use of definition (41) in the text, Eq.\ (\ref{SM2:16}) is expressed as
\begin{equation}
\frac{{\rm d}L(\sqrt{{\tilde \eta}w/K})}{{\rm d}w}\simeq
-\frac{2{\tilde \eta}}{{\tilde C}^{*}\sqrt{K}}\frac{2}{{\tilde B}^{*2}}H({\tilde B}^{\,*}).
\label{SM2:16a}
\end{equation}
In the limit of $w\to \infty$, the following asymptotic formulas hold: 
\begin{equation}
\log\left|
\frac{\sqrt{w+1}-\sqrt{w}}{\sqrt{w+1}+\sqrt{w}}\right|\simeq -\log|4w|,
\label{SM2:17}
\end{equation}
and 
\begin{equation}
\sqrt{w(w+1)}-w\simeq \frac{1}{{2}}.
\label{SM2:18}
\end{equation}
Therefore, $X^{(2)}(\eta)$, Eq.\ (\ref{SM2:14}), is given by
\begin{eqnarray}
& &X^{(2)}(\eta)\simeq
\frac{{\tilde \eta}}{2K^{\,3/2}}
\frac{2{\tilde \eta}}{{\tilde C}^{*}\sqrt{K}}
{\frac{2}{{\tilde B}^{*2}}}
H({\tilde B}^{\,*})
\int^{2K/{\tilde \eta}}{\rm d}w
\left({-\frac{1}{2}}-\frac{1}{2}\log|4w|\right)
\nonumber
\\
& &
\qquad
\simeq\frac{1}{4}\frac{\tilde \eta}{K^{\,3/2}}\log\frac{e^{\,{2}}\,{\tilde \eta}}{8K}\times
\frac{4}{{\tilde B}^{\,*}}H({\tilde B}^{\,*}).
\label{SM2:19}
\end{eqnarray}
In deriving the last approximate result in Eq.\ (\ref{SM2:19}), we have discarded the term of 
the order of ${\cal O}(\eta^{2}\log\eta)$, and used the definition of ${\tilde B}^{\,*}$, 
Eq.\ (\ref{SM2:13}) .  

Collecting the above results, Eqs.\ (\ref{SM2:11}), (\ref{SM2:12a}), and (\ref{SM2:19}), 
near $\eta\sim0$, $X(\eta)$, Eq.\ (\ref{SM2:10}), is given by 
\begin{equation}
X(\eta)\simeq\frac{1}{4}\frac{\tilde \eta}{K^{\,3/2}}
\frac{4}{{\tilde C}^{*}{\tilde B}^{\,*}}\left[G(2{\tilde B}^{\,*})\,{\log\frac{e\,{\tilde \eta}}{8K}}
+H({\tilde B}^{\,*})\,{\log\frac{e^{\,2}\,{\tilde \eta}}{8K}}\right].
\label{SM2:20}
\end{equation}
With the use of definitions of ${\tilde B}^{\,*}$, Eq.\ (\ref{SM2:13}), 
${\tilde C}^{*}\equiv {\tilde C}v_{\rm F}^{\,2}/A$, and 
${\tilde \eta}\equiv v_{\rm F}^{\,2}\eta/A$, Eq.\ (\ref{SM2:20}) is transformed into 
a compact form as 
\begin{equation}
X(\eta)\simeq\frac{1}{4K}\frac{\eta}{Ak_{\rm F}^{\,2}}
\left[G\left(\frac{4\sqrt{2}m^{*}A}{{\tilde C}}\right)\,{\log\frac{e\,\eta}{16Ak_{\rm F}^{\,2}}}
+H\left(\frac{2\sqrt{2}m^{*}A}{{\tilde C}}\right)\,{\log\frac{e^{\,2}\,\eta}{16Ak_{\rm F}^{\,2}}}\right].
\label{SM2:21}
\end{equation}
Then, $S_{3{\rm dF}}(\eta)-S_{3{\rm dF}}(0)$, Eq.\ (\ref{SM2:7}), is given by
\begin{equation}
S_{3{\rm dF}}(\eta)-S_{3{\rm dF}}(0)
\simeq\frac{1}{4K}\frac{\eta}{Ak_{\rm F}^{\,2}}
\left[G\left(\frac{4\sqrt{2}m^{*}A}{{\tilde C}}\right)\,{\log\frac{e\,\eta}{16Ak_{\rm F}^{\,2}}}
+H\left(\frac{2\sqrt{2}m^{*}A}{{\tilde C}}\right)\,{\log\frac{e^{\,2}\,\eta}{16Ak_{\rm F}^{\,2}}}\right].
\label{SM2:22}
\end{equation}

Finally, adding Eq.\ (\ref{SM2:6}) and Eq.\ (\ref{SM2:22}), the second term of Eq.\ (37) in the text 
is reduced to the first and second terms in the brace of Eq.\ (38) in the text.  

{
\section{Derivation of Eq.\ (42) in the text}
Here we derive Eq.\ (42) in the text for the case of 3D-FQCP.  
We start with Eq.\ (\ref{eq:G01}),  
the same equation as that in the case of 3D-AFQCP except that $C^{*}$, in the parameter 
${\tilde a}_{\pm}\equiv a+C^{*}|\pm\varepsilon_{n}-\varepsilon_{n^{\,\prime}}|$, is not constant 
but $C^{*}\equiv {\tilde C}^{*}/\sqrt{1-\cos\,\theta}$ with ${\tilde C}^{*}$ being a constant. 
Then, we  obtain the same expression for ${\bar \Gamma}_{\nu}\ks{\mathbf{k},\rm{i}\varepsilon_{n}}$ 
as Eq.\ (\ref{eq:G02}).   
}
{
However, since Eq.\ (\ref{eq:G02}) does not include the coefficient $C^{*}$, the expression 
Eq.\ (\ref{eq:G04}) for ${\bar \Gamma}_{\nu}\ks{\mathbf{k},\rm{i}\varepsilon_{n}}$ can be also 
applied to the present case of 3D-FQCP:
\begin{eqnarray}
& &
{\bar \Gamma}_{\nu}\ks{\mathbf{k},\rm{i}\varepsilon_{n}}
=J\gamma_{\nu}({\bf k})\,\pi T\,
\,\int_{-1}^{1}\frac{{\rm d}x}{2}
\left\{\frac{1}{[{\tilde \eta}+K(1-x)]^{3/2}}-
\frac{2\varepsilon_{n}}{[{\tilde \eta}+K(1-x)]^{2}}
\right.
\nonumber
\\
& &
\left.
\qquad\qquad\qquad\qquad\qquad
+2{\rm i}\sqrt{\frac{K}{2}}\frac{(1-x)}{[{\tilde \eta}+K(1-x)]^{2}}
-6{\rm i}\sqrt{\frac{K}{2}}\frac{\varepsilon_{n}}{[{\tilde \eta}+K(1-x)]^{5/2}}
\right\}.
\label{eq:G043dF}
\end{eqnarray}

Then, the expression Eq.\ (\ref{eq:G06}) for 
${\bar \Gamma}_{\nu}\ks{\mathbf{k},\rm{i}\varepsilon_{n}}$ is also valid in the case of 3d-FQCP:
\begin{equation}
{\bar \Gamma}_{\nu}\ks{\mathbf{k},\rm{i}\varepsilon_{n}}
\simeq
J\gamma_{\nu}({\bf k})\,
\left[
\frac{\sqrt{Ak_{\rm F}^{2}}}{2(k_{\rm F}v_{\rm F})^{3}}\frac{\pi T}{\sqrt{\eta}}
-\,\varepsilon_{n}\,\frac{Ak_{\rm F}^{2}}{2(k_{\rm F}v_{\rm F})^{4}}
\frac{\pi T}{\eta}
+\frac{{\rm i}\,\pi T}{4(k_{\rm F}v_{\rm F})^{3}}\,\log\frac{4Ak_{\rm F}^{2}}{\eta}
-{\rm i}\varepsilon_{n}\,\frac{\sqrt{Ak_{\rm F}^{2}}}{(k_{\rm F}v_{\rm F})^{4}}\,
\frac{\pi T}{\sqrt{\eta}}
\right].
\label{eq:G063dF}
\end{equation}

Since $\eta\propto T^{\,4/3}$ just at the 3D-FQCP,~\cite{MoriyaTakimoto} 
the first and fourth terms in Eq.\ (\ref{eq:G063dF}) vanishes as $T^{1/3}$ 
while the second term diverges as $T^{-1/3}$.  
Although the first term in Eq.\ (\ref{eq:G063dF}) vanishes as $T^{1/3}$ in the limit $T\to 0$ at the 
criticality, this $T$ dependence gives a sharper cusp than the second term of Eq.\ (39) in the text 
because $\eta\,\log\,\eta\propto T^{4/3}\log T$ there.  
}

{
The summation over $\varepsilon_{n^{\,\prime}}$ (except for the term 
$\varepsilon_{n^{\,\prime}}=\varepsilon_{n}$) in Eq.\ (\ref{eq:G01}) 
can be approximated, in the region $T\to 0$, by 
integration with respect to $\varepsilon^{\prime}$ as discussed in Sect.\ 3.  
Since the final results for this 
contribution in the case of 3D-AFQCP, i.e.,
}
{Eqs.\ (\ref{eq:G6b}) $\sim$ (\ref{eq:G7b}), also do not include 
the parameter $C^{*}$, they are also applied to the case of 3D-FQCP. 

Then, $\Gamma_{\nu}\ks{\mathbf{k},\rm{i}\varepsilon_{n}}$ is given by the same expression as 
Eq.\ (\ref{eq:G8}): 
\begin{equation}
\Gamma_{\nu}\ks{\mathbf{k},\rm{i}\varepsilon_{n}}
\simeq
\Gamma_{\nu}\ks{\mathbf{k},\rm{i}\varepsilon_{n}}|_{\varepsilon_{n}=0}
+J\gamma_{\nu}({\bf k}-{\bf Q})\left[
-\frac{\sqrt{Ak_{\rm F}^{2}}}{(k_{\rm F}v_{\rm F})^{4}}\frac{\pi T}{\sqrt{\eta}}
+\frac{1}{4(k_{\rm F}v_{\rm F})^{3}}\left(\log\,\frac{4Ak_{\rm F}^{2}}{\eta e}\right)
\right]({\rm i}\varepsilon_{n})
+{\cal O}(\varepsilon_{n}^{2}).
\label{eq:G8F}
\end{equation}
}
{
This gives Eq.\ (42) in the text.  
}

{
As in the case of 3D-AFQCP, the term $-2{\rm i}b\,\pi T/a^{2}$ in Eq.\ (\ref{eq:G6})  
cancels out the third term in Eqs.\ (\ref{eq:G043dF}) and (\ref{eq:G063dF}).
}

\section{Derivation of Eq.\ (43) in the text}
Near the 2D-AFQCP, instead of Eq.\ (24) in the text, 
the correction to ASSO coupling $\alpha_{\rm af}(\eta)$ is given by 
\begin{equation}
\alpha_{\rm af}(\eta)=
\frac{{\bar \Gamma}_{\nu}\ks{\mathbf{k},{\rm i}\delta}}{\gamma_{\nu}({\bf k}-{\bf Q})}+
\frac{2J}{C^{*}}\int_{0}^{2\pi}\frac{{\rm d}\varphi}{2\pi}
\frac{1}{\sqrt{a^{*}}}\int_{1}^{\infty}{\rm d}y\,
\frac{1}{\left[y+\frac{\sqrt{a^{*}}}{C^{*}}(y^{\,2}-1)\right]^{2}},
\label{SM3:1}
\end{equation} 
where $\cos\,\varphi\equiv({\bf k}\cdot{\bf k}^{\prime})/(|{\bf k}||{\bf k}^{\prime}|)$, and 
\begin{equation}
a^{*}\equiv {\tilde \eta}+K(1-\cos\,\varphi).
\label{SM3:2}
\end{equation}
Let us introduce ${\tilde \alpha}_{\rm af}(\eta)$ expressing the second term on the r.h.s. of 
Eq.\ (\ref{SM3:1}).  
Since the leading singularity for ${\tilde \alpha}_{\rm af}(\eta)$ in $\eta$ arises from the integration 
with respect to $\varphi$ 
near $\cos\,\varphi\sim 1$ of the term $1/\sqrt{a^{*}}$, in the denominator of 
integration with respect to $y$ can be approximated by putting $a^{*}\to 0$: 
\begin{eqnarray}
& &
{\tilde \alpha}_{\rm af}(\eta)\simeq
\frac{2J}{C^{*}}\int_{0}^{2\pi}\frac{{\rm d}\varphi}{2\pi}
\frac{1}{\sqrt{{\tilde \eta}+K(1-\cos\varphi)}}
\int_{1}^{\infty}{\rm d}y\,
\frac{1}{y^{\,2}}
\nonumber
\\
& &
\qquad
=\frac{2J}{C^{*}\sqrt{K}}
\int_{0}^{2\pi}\frac{{\rm d}\varphi}{2\pi}
\frac{1}{\sqrt{1+({\tilde \eta}/K)-\cos\varphi}}.
\label{SM3:3}
\end{eqnarray} 
The integration with respect to $\varphi$ in Eq.\ (\ref{SM3:3}) is given by 
$2F\left(\pi/2,1/\sqrt{1+({\tilde \eta}/2K})\right)/\pi\sqrt{2+({\tilde \eta}/K)}$, 
where $F(\pi/2,x)=K(x)$ is the complete elliptic function of first kind.  
The approximate form of $K(x)$ near $x \simeq 1$ is given by  
\begin{equation}
K(x)\simeq\frac{1}{2}\log\frac{16}{1-x^{\,2}}.
\label{SM3:4}
\end{equation}
Therefore, in the limit of $\eta\to 0$, we obtain 
\begin{equation}
{\tilde \alpha}_{\rm af}(\eta)\simeq
\frac{\sqrt{2}J}{\pi C^{*}\sqrt{K}}\log\frac{32K}{{\tilde \eta}}.
\label{SM3:5}
\end{equation}
Substituting definitions, $C^{*}\equiv v_{\rm F}^{\,2}C/A$, $K\equiv 2k_{\rm F}^{\,2}v_{\rm F}^{\,2}$, 
and ${\tilde \eta}\equiv v_{\rm F}^{\,2}\eta/A$, into Eq.\ (\ref{SM3:5}), we obtain
\begin{equation}
{\tilde \alpha}_{\rm af}(\eta)\simeq
\frac{J}{2k_{\rm F}^{\,2}v_{\rm F}^{\,2}}\frac{2m^{*}A}{\pi v_{\rm F}C}
\log\frac{64Ak_{\rm F}^{\,2}}{\eta}.
\label{SM3:6}
\end{equation}

Finally, by shifting $\eta$ to $\eta+Aq_{m}^{\,2}$ to take into account the degree of distance 
from the hot point, we obtain 
the first term in the parenthesis on the r.h.s. of Eq.\ (43) in the text.

\section{Derivation of Eq.\ (44) in the text}
Near the 2D-FQCP, instead of Eq.\ (37) in the text, 
the correction to ASSO coupling $\alpha_{\rm f}(\eta)$ is given by 
\begin{eqnarray}
& &
\alpha_{\rm f}(\eta)\equiv 
\frac{{\bar \Gamma}_{\nu}\ks{\mathbf{k},{\rm i}\delta}}{\gamma_{\nu}({\bf k})}+
2J\int_{0}^{2\pi}\frac{{\rm d}\varphi}{2\pi}
\frac{\sqrt{1-\cos\varphi}}{{\tilde C}^{*}\sqrt{a^{*}}}\int_{1}^{\infty}{\rm d}y\,
\frac{1}{\left[y+\frac{\sqrt{a^{*}(1-\cos\varphi)}}{{\tilde C}^{*}}(y^{\,2}-1)\right]^{2}}
\nonumber
\\
& &
\qquad
\equiv 
\frac{{\bar \Gamma}_{\nu}\ks{\mathbf{k},{\rm i}\delta}}{\gamma_{\nu}({\bf k})}+ 
2JS_{2{\rm dF}}(\eta),
\label{SM4:1}
\end{eqnarray} 
where ${\tilde C}^{*}\equiv v_{\rm F}^{\,2}{\tilde C}/A$.   
Substituting Eq.\ (\ref{SM3:2}) into the integrands of the second term on r.h.s. of 
Eq.\ (\ref{SM4:1}), $S_{2{\rm dF}}(\eta)$ is expressed as 
\begin{equation}
S_{2{\rm dF}}(\eta)=\int_{0}^{2\pi}\frac{{\rm d}\varphi}{2\pi}
\frac{\sqrt{1-\cos\varphi}}{{\tilde C}^{*}\sqrt{{\tilde \eta}+K(1-\cos\varphi)}}
\int_{1}^{\infty}{\rm d}y\,
\frac{1}{\left[y+
\frac{\sqrt{1-\cos\varphi}\sqrt{{\tilde \eta}+K(1-\cos\varphi)}}{{\tilde C}^{*}}(y^{\,2}-1)\right]^{2}}.
\label{SM4:2}
\end{equation}

First, we calculate $S_{2{\rm dF}}(0)$ which is considerably simplified: 
\begin{equation}
S_{2{\rm dF}}(0)=
\frac{1}{{\tilde C}^{*}\sqrt{K}}
\int_{1}^{\infty}{\rm d}y\,
\int_{0}^{2\pi}\frac{{\rm d}\varphi}{2\pi}
\frac{1}{\left[y+
\frac{\sqrt{K}}{{\tilde C}^{*}}(1-\cos\varphi)(y^{\,2}-1)\right]^{2}},
\label{SM4:3}
\end{equation} 
where we have interchanged the order of the integrations with respect to $\varphi$ and $y$.   
The integration with respect to $\varphi$  is easily performed to give 
\begin{equation}
S_{2{\rm dF}}(0)=
\frac{1}{{\tilde C}^{*}\sqrt{K}}
\int_{1}^{\infty}{\rm d}y\,
\frac{1}{y\,^{3/2}}\frac{y+\frac{\sqrt{K}}{{\tilde C}^{*}}(y^{\,2}-1)}
{\left[y+\frac{2\sqrt{K}}{{\tilde C}^{*}}(y^{\,2}-1)\right]^{3/2}}.
\label{SM4:4}
\end{equation} 
With the use of a function $I(x)$, defined by Eq.\ (45) in the text, $S_{2{\rm dF}}(0)$ is 
expressed as 
\begin{equation}
S_{2{\rm dF}}(0)=\frac{1}{K}I\left(\frac{\sqrt{K}}{{\tilde C}^{*}}\right).
\label{SM4:5}
\end{equation} 

Next, let us discuss the lowest-order correction in $\eta$, which will turn out to be of 
the order of ${\cal O}(\sqrt{\eta})$.  To this end, first, we change the integration variable 
in Eq.\ (\ref{SM4:2}) from $\varphi$ to $x\equiv\cos(\varphi/2)$: 
\begin{equation}
S_{2{\rm dF}}(\eta)=
\frac{\sqrt{2}}{\pi{\tilde C}^{*}}
\int_{-1}^{1}{\rm d}x
\frac{1}{\sqrt{{\tilde \eta}+2K(1-x^{\,2})}}
\int_{1}^{\infty}{\rm d}y\,
\frac{1}{\left[y+
\frac{\sqrt{2}}{{\tilde C}^{*}}\sqrt{1-x^{\,2}}\sqrt{{\tilde \eta}+2K(1-x^{\,2})}\,(y^{\,2}-1)\right]^{2}}.
\label{SM4:6}
\end{equation}
Let us define the integration with respect to $y$ in Eq.\ (\ref{SM4:6}) as $M(x,{\tilde \eta})$: i.e., 
\begin{equation}
M(x,{\tilde \eta})\equiv
\int_{1}^{\infty}{\rm d}y\,
\frac{1}{\left[y+
\frac{\sqrt{2}}{{\tilde C}^{*}}\sqrt{1-x^{\,2}}\sqrt{{\tilde \eta}+2K(1-x^{\,2})}\,(y^{\,2}-1)\right]^{2}}.
\label{SM4:7}
\end{equation}
Then, 
\begin{equation}
S_{2{\rm dF}}(\eta)-S_{2{\rm dF}}(0)=\frac{2}{\pi{\tilde C}^{*}\sqrt{K}}
\int_{0}^{1}{\rm d}x
\left[\frac{1}{\sqrt{1+({\tilde \eta}/2{K})-x^{\,2}}}M(x,{\tilde \eta})
-\frac{1}{\sqrt{1-x^{\,2}}}M(x,0)
\right],
\label{SM4:8}
\end{equation}
where we have used the fact that $M(x,{\tilde \eta})$ is an even function in $x$, and 
changed the interval of the integration with respect to $x$ from $-1\le x \le 1$ to $0\le x \le 1$, 
multiplying the result by a factor 2.  The expression (\ref{SM4:8}) is rearranged as 
\begin{eqnarray}
& &
S_{2{\rm dF}}(\eta)-S_{2{\rm dF}}(0)=\frac{2}{\pi{\tilde C}^{*}\sqrt{K}}
\left\{
\int_{0}^{1}{\rm d}x
\left[\frac{1}{\sqrt{1+({\tilde \eta}/2{K})-x^{\,2}}}
-\frac{1}{\sqrt{1-x^{\,2}}}\right]M(x,{\tilde \eta})
\right.
\nonumber
\\
& &
\left.
\qquad\qquad\qquad\qquad\qquad\qquad
+\int_{0}^{1}{\rm d}x
\frac{1}{\sqrt{1-x^{\,2}}}
\left[M(x,{\tilde \eta})-M(x,0)\right]
\right\}.
\label{SM4:9}
\end{eqnarray}
Let us introduce $Y^{(1)}(\eta)$ representing the first term in the brace of Eq.\ (\ref{SM4:9}).  
By making partial integration with respect to $x$, $Y^{(1)}(\eta)$ is expressed as 
\begin{eqnarray}
& &
Y^{\,(1)}(\eta)=
\left[\sin^{-1}\frac{1}{\sqrt{1+({\tilde \eta}/2{K})}}
-\sin^{-1}1\right]M(1,{\tilde \eta})
\nonumber
\\
& &
\qquad\qquad\qquad
-\int_{0}^{1}{\rm d}x
\left[\sin^{-1}\frac{x}{\sqrt{1+({\tilde \eta}/2{K})}}
-\sin^{-1}x\right]\frac{\partial M(x,{\tilde \eta})}{\partial x}.
\label{SM4:10}
\end{eqnarray}
In the limit of $\eta\to 0$, Eq.\ (\ref{SM4:10}) is reduced to 
\begin{equation}
Y^{\,(1)}(\eta)\simeq
{-}\sqrt{\frac{{\tilde \eta}}{2K}}M(1,0)
-\int_{0}^{1}{\rm d}x
\left[-\frac{1}{2}\frac{x}{\sqrt{1-x^{\,2}}}\frac{{\tilde \eta}}{2{K}}+{\cal O}({\tilde \eta}^{2})
\right]
\frac{\partial M(x,0)}{\partial x}.
\label{SM4:11}
\end{equation}
By straightforward calculations, $M(1,0)$ and ${\partial M(x,0)}/{\partial x}$ are reduced to 
\begin{equation}
M(1,0)=\int_{{1}}^{\infty}{\rm d}y\,\frac{1}{y^{\,2}}=1,
\label{SM4:12}
\end{equation}
and
\begin{equation}
\frac{\partial M(x,0)}{\partial x}=\frac{8\sqrt{K}}{{\tilde C}^{*}}x\int_{1}^{\infty}
{\rm d}y\,\frac{y^{\,2}-1}{\left[y+\frac{2\sqrt{K}}{{\tilde C}^{*}}(1-x^{\,2})(y^{\,2}-1)\right]^{3}}.
\label{SM4:13}
\end{equation}
Namely, up to the leading order in $\eta$, $Y^{\,(1)}(\eta)$ is given by 
\begin{equation}
Y^{\,(1)}(\eta)\simeq -\sqrt{\frac{{\tilde \eta}}{2K}}+{\cal O}({\tilde \eta}).
\label{SM4:14}
\end{equation}
It is easy to see that the second term in the brace of Eq.\ (\ref{SM4:9}) gives only 
the term of the order of ${\cal O}({\tilde \eta})$ at most.  
Therefore, $S_{2{\rm dF}}(\eta)-S_{2{\rm dF}}(0)$, Eq.\ (\ref{SM4:9}), is given by
\begin{eqnarray}
& & 
S_{2{\rm dF}}(\eta)-S_{2{\rm dF}}(0)
\simeq
-\frac{2}{\pi\sqrt{2}{\tilde C}^{*}K}\sqrt{{\tilde \eta}}+{\cal O}({\tilde \eta})
\nonumber
\\
& &
\qquad\qquad\qquad\qquad\qquad
=-\frac{1}{K}\frac{\sqrt{2A}}{\pi{\tilde C}v_{\rm F}}\sqrt{\eta}+{\cal O}(\eta),
\label{SM4:15}
\end{eqnarray}
where we have used the definitions, ${\tilde C}^{*}\equiv {\tilde C}v_{\rm F}^{\,2}/A$ 
and ${\tilde \eta}\equiv v_{\rm F}^{\,2}\eta/A$, in deriving the last equality.  

Finally, adding Eq.\ (\ref{SM4:5}) and Eq.\ (\ref{SM4:15}) and using the definition 
$K\equiv 2k_{\rm F}^{\,2}v_{\rm F}^{\,2}$, $\alpha_{\rm f}(\eta)$, Eq.\ (\ref{SM4:1}), is given by 
\begin{equation}
\alpha_{\rm f}(\eta)=
\frac{{\bar \Gamma}_{\nu}\ks{\mathbf{k},{\rm i}\delta}}{\gamma_{\nu}({\bf k})}+
\frac{J}{2k_{\rm F}^{\,2}v_{\rm F}^{\,2}}
\left[2I\left(\frac{\sqrt{K}}{{\tilde C}^{*}}\right)-\frac{2\sqrt{2A}}
{\pi{\tilde C}v_{\rm F}}\sqrt{\eta}
+{\cal O}(\eta)
\right].
\label{SM4:16}
\end{equation}
The second term on the r.h.s. of Eq.\ (\ref{SM4:16}) is nothing but the first and 
second terms in the bracket on the r.h.s. of 
Eq.\ (44) in the text because 
$\sqrt{K}/{\tilde C}^{*}=\sqrt{2}m^{*}A/{\tilde C}$.  

{
\section{Derivation of Eq.\ (46) in the text for 2D-AFQCP}
Here we derive Eq.\ (46) in the text for 2D-AFQCP. Instead of Eq.\ (\ref{eq:G01}), 
$\Gamma_{\nu}\ks{\mathbf{k},\rm{i}\varepsilon_{n}}$ in 2D is given by 
\begin{eqnarray}
& &
\Gamma_{\nu}\ks{\mathbf{k},\rm{i}\varepsilon_{n}}
=J\gamma_{\nu}({\bf k}-{\bf Q})\,\int_{0}^{2\pi}\frac{{\rm d}\varphi}{2\pi}
\nonumber
\\
& &\qquad\qquad\qquad\qquad
\pi 
T\sum_{n^{\,\prime}\ge 0}\left\{
\frac{1}{\sqrt{{\tilde a}_{+}}}
\frac{1}{(\sqrt{{\tilde a}_{+}}+\varepsilon_{n^{\,\prime}})^{2}}
+
\frac{1}{\sqrt{{\tilde a}_{-}}}
\frac{1}{(\sqrt{{\tilde a}_{-}}+\varepsilon_{n^{\,\prime}})^{2}}
\right.
\nonumber
\\
& &
\left.
\qquad\qquad\qquad\qquad\qquad
+2{\rm i}b
\left[
\frac{1}{\sqrt{{\tilde a}_{+}}}
\frac{1}{(\sqrt{{\tilde a}_{+}}+\varepsilon_{n^{\,\prime}})^{3}}
-
\frac{1}{\sqrt{{\tilde a}_{-}}}
\frac{1}{(\sqrt{{\tilde a}_{-}}+\varepsilon_{n^{\,\prime}})^{3}}
\right]
\right\}, 
\label{eq:2dG01}
\end{eqnarray} 
where ${\tilde a}_{\pm}\equiv a+C^{*}|\pm\varepsilon_{n}-\varepsilon_{n^{\,\prime}}|$ with 
$a$ given by Eq.\ (15) in the text as 
$a=(v_{\rm F}^{\,2}\eta/A)+2k_{\rm F}^{\,2}v_{\rm F}^{\,2}(1-\cos\,\varphi)$, and 
$b\equiv k_{\rm F}v_{\rm F}(1-\cos\,\varphi)$.  
}

{
Corresponding to Eq.\ (\ref{eq:G02}), the contribution 
from $\varepsilon_{n\,^{\prime}}=\varepsilon_{n}$ in the summation over $\varepsilon_{n\,^{\prime}}$ 
in 
}
{
the first and the third terms in the brace of Eq.\ (\ref{eq:2dG01}) is given by
\begin{equation}
{\bar \Gamma}_{\nu}\ks{\mathbf{k},\rm{i}\varepsilon_{n}}
=J\gamma_{\nu}({\bf k}-{\bf Q})\,\pi T\,\int_{0}^{2\pi}\frac{{\rm d}\varphi}{2\pi}
\left[
\frac{1}{\sqrt{a}}\frac{1}{(\sqrt{a}+\varepsilon_{n})^{\,2}}
+2{\rm i}b\,
\frac{1}{\sqrt{a}}
\frac{1}{(\sqrt{a}+\varepsilon_{n})^{\,3}}
\right].
\label{eq:2dG02}
\end{equation} 
}
{
Expanding the expression in the bracket up to linear order in $\varepsilon_{n}$ as in 
Eq.\ (\ref{eq:G03}), Eq.\ (\ref{eq:2dG02}) is reduced to  
}
{
\begin{eqnarray}
& &
{\bar \Gamma}_{\nu}\ks{\mathbf{k},\rm{i}\varepsilon_{n}}
=J\gamma_{\nu}({\bf k}-{\bf Q})\,\pi T\int_{0}^{2\pi}\frac{{\rm d}\varphi}{2\pi}
\left\{\frac{1}{[{\tilde \eta}+K(1-\cos\,\varphi)]^{3/2}}-
\frac{2\varepsilon_{n}}{[{\tilde \eta}+K(1-\cos\,\varphi)]^{2}}
\right.
\nonumber
\\
& &
\left.
\qquad\qquad\qquad\qquad
+2{\rm i}\sqrt{\frac{K}{2}}\frac{(1-\cos\,\varphi)}
{[{\tilde \eta}+K(1-\cos\,\varphi)]^{2}}
-6{\rm i}\sqrt{\frac{K}{2}}\frac{(1-\cos\,\varphi)\varepsilon_{n}}
{[{\tilde \eta}+K(1-\cos\,\varphi)]^{5/2}}
\right\}.
\label{eq:2dG04}
\end{eqnarray}
}

{
The integration with respect to $\varphi$ of the first term in Eq.\ (\ref{eq:2dG04}) is performed 
as follows:
\begin{eqnarray}
& &
\int_{0}^{2\pi}\frac{{\rm d}\varphi}{2\pi}
\frac{1}{[{\tilde \eta}+K(1-\cos\,\varphi)]^{3/2}}
\nonumber
\\
& &
\quad\qquad
=(-2)\frac{\partial}{\partial {\tilde \eta}}\int_{0}^{2\pi}\frac{{\rm d}\varphi}{2\pi}
\frac{1}{\sqrt{{\tilde \eta}+K(1-\cos\,\varphi)}}
\nonumber
\\
& &
\quad\qquad
=-\frac{2}{\sqrt{K}}\frac{\partial}{\partial {\tilde \eta}}
\int_{0}^{2\pi}\frac{{\rm d}\varphi}{2\pi}
\frac{1}{\sqrt{1+({\tilde \eta}/K)-\cos\,\varphi}}.
\label{eq:2dG06}
\end{eqnarray}
The integration with respect to $\varphi$ in the last line of Eq.\ (\ref{eq:2dG06}) is the same 
as that appeared 
in Eq.\ (\ref{SM3:3}) and is given by 
$2F\left(\pi/2,1/\sqrt{1+({\tilde \eta}/2K})\right)/\pi\sqrt{2+({\tilde \eta}/K)}$, 
where $F(\pi/2,x)=K(x)$ is the complete elliptic function of first kind and its 
approximate form near $x \simeq 1$ is given by Eq.\ (\ref{SM3:4}).  Therefore, in the limit 
${\tilde \eta}\to 0$, 
\begin{equation}
\int_{0}^{2\pi}\frac{{\rm d}\varphi}{2\pi}
\frac{1}{\sqrt{1+({\tilde \eta}/K)-\cos\,\varphi}}\simeq\frac{1}{\pi\sqrt{2}}
\log\,\frac{32}{({\tilde \eta}/K)}.
\label{eq:2dG07}
\end{equation}
Substituting this asymptotic form into the last line of Eq.\ (\ref{eq:2dG06}), we obtain 
\begin{equation}
\int_{0}^{2\pi}\frac{{\rm d}\varphi}{2\pi}
\frac{1}{[{\tilde \eta}+K(1-\cos\,\varphi)]^{3/2}}
\simeq
\frac{1}{\sqrt{K}}\frac{\sqrt{2}}{\pi}\frac{1}{{\tilde \eta}}.
\label{eq:2dG08}
\end{equation}
}

{
The integration with respect to $\varphi$ of the third term in Eq.\ (\ref{eq:2dG04}) is 
performed as follows:
\begin{eqnarray}
& &
\int_{0}^{2\pi}\frac{{\rm d}\varphi}{2\pi}
\frac{(1-\cos\,\varphi)}{[{\tilde \eta}+K(1-\cos\,\varphi)]^{5/2}}
\nonumber
\\
& &
\quad\qquad
=\left(-\frac{2}{3}\right)\frac{\partial}{\partial K}\int_{0}^{2\pi}\frac{{\rm d}\varphi}{2\pi}
\frac{1}{[{\tilde \eta}+K(1-\cos\,\varphi)]^{3/2}}.
\label{eq:2dG09}
\end{eqnarray}
Substituting the relation (\ref{eq:2dG08}) into Eq.\ (\ref{eq:2dG09}), we obtain
\begin{equation}
\int_{0}^{2\pi}\frac{{\rm d}\varphi}{2\pi}
\frac{(1-\cos\,\varphi)}{[{\tilde \eta}+K(1-\cos\,\varphi)]^{5/2}}
\simeq
\frac{1}{K^{3/2}}\frac{\sqrt{2}}{3\pi}\frac{1}{{\tilde \eta}}.
\label{eq:2dG010}
\end{equation}

The integration with respect to $\varphi$ of the last term in Eq.\ (\ref{eq:2dG04}) is performed as follows: 
\begin{equation}
\int_{0}^{2\pi}\frac{{\rm d}\varphi}{2\pi}
\frac{1}{[{\tilde \eta}+K(1-\cos\,\varphi)]^{2}}
=\frac{K+{\tilde \eta}}{[(K+{\tilde \eta})^{2}-K^{2}]^{3/2}}.
\label{eq:2dG0611}
\end{equation}
Then, to the leading order in $1/{\tilde \eta}\gg 1$, we obtain
\begin{equation}
\int_{0}^{2\pi}\frac{{\rm d}\varphi}{2\pi}
\frac{1}{[{\tilde \eta}+K(1-\cos\,\varphi)]^{2}}
\simeq
\frac{K}{(2K{\tilde \eta})^{3/2}}.
\label{eq:2dG012}
\end{equation}

The integration with respect to $\varphi$ 
of the third term in the bracket of Eq.\ (\ref{eq:2dG04}) {is performed as follows:
\begin{eqnarray}
& &
\int_{0}^{2\pi}\frac{{\rm d}\varphi}{2\pi}
\frac{b}{a^{2}}=
\sqrt{\frac{K}{2}}\,
\int_{0}^{2\pi}\frac{{\rm d}\varphi}{2\pi}
\frac{(1-\cos\,\varphi)}{[{\tilde \eta}+K(1-\cos\,\varphi)]^{2}}
\nonumber
\\
& &
\qquad\qquad
=\sqrt{\frac{K}{2}}\,\left[-\frac{\partial}{\partial K}
\int_{0}^{2\pi}\frac{{\rm d}\varphi}{2\pi}
\frac{1}{{\tilde \eta}+K(1-\cos\,\varphi)}\right]
=\sqrt{\frac{K}{2}}\,\left[-\frac{\partial}{\partial K}\,
\frac{1}{\sqrt{(K+{\tilde \eta})^{2}-K^{2}}}\right]
\nonumber
\\
& &
\qquad\qquad
\simeq
\sqrt{\frac{K}{2}}\,\left[-\frac{\partial}{\partial K}\,
\frac{1}{\sqrt{2K{\tilde \eta}}}\right]
=\frac{1}{4K}\frac{1}{\sqrt{{\tilde \eta}}}.
\label{eq:2dG8}
\end{eqnarray}
}

Substituting Eqs.\ (\ref{eq:2dG08}), (\ref{eq:2dG010}), (\ref{eq:2dG012}), and 
(\ref{eq:2dG8}) into Eq.\ (\ref{eq:2dG04}), we obtain  
\begin{equation}
{\bar \Gamma}_{\nu}\ks{\mathbf{k},\rm{i}\varepsilon_{n}}\simeq
J\gamma_{\nu}({\bf k}-{\bf Q})\,\frac{T}{\sqrt{K}}
\left(\sqrt{2}\frac{1}{{\tilde \eta}}
-\frac{\pi}{\sqrt{2}}\frac{\varepsilon_{n}}{{\tilde \eta}^{3/2}}
+{\rm i}\frac{\pi}{2\sqrt{K}}\frac{1}{\sqrt{{\tilde \eta}}}
-\frac{2{\rm i}}{\sqrt{K}}\,
\frac{\varepsilon_{n}}{{\tilde \eta}}
\right).
\label{eq:2dG013}
\end{equation}
Substituting ${\tilde \eta}\equiv v_{\rm F}^{\,2}\eta/A$ 
and $K\equiv 2k_{\rm F}^{\,2}v_{\rm F}^{\,2}$ into this expression, we obtain
\begin{equation}
{\bar \Gamma}_{\nu}\ks{\mathbf{k},\rm{i}\varepsilon_{n}}
\simeq
J\gamma_{\nu}({\bf k}-{\bf Q})\,
\left[
\frac{Ak_{\rm F}^{2}}{(k_{\rm F}v_{\rm F})^{3}}\frac{T}{\eta}
-\,\varepsilon_{n}\,\frac{(Ak_{\rm F}^{2})^{3/2}}{2(k_{\rm F}v_{\rm F})^{4}}
\frac{\pi T}{\eta^{3/2}}
+{\rm i}\frac{\sqrt{Ak_{\rm F}^{2}}}{4(k_{\rm F}v_{\rm F})^{3}}
\frac{\pi T}{\sqrt{\eta}}
-{\rm i}\varepsilon_{n}\,\frac{Ak_{\rm F}^{2}}{(k_{\rm F}v_{\rm F})^{4}}
\frac{T}{\eta}
\right].
\label{eq:2dG014}
\end{equation}
Since $\eta\propto T/(-\log\,T)$ in the case of 2D-AFQCP,~\cite{Moriya2} 
the first and the fourth terms in Eq.\ (\ref{eq:2dG014}) diverge, in the limit $T\to 0$, 
as $(-\log\,T)$, while the second term diverges as $(-\log\,T)^{3/2}/T^{1/2}$.  
We remark here that the first term in Eq.\ (\ref{eq:2dG014}) should be included in Eq.\ (43) 
in the text because it gives the same divergence in the $T$ dependence as the first term 
in Eq.\ (43) in the text
at least on the hot line (i.e., $q_{m}=0$) in the limit $T\to 0$.   
}

{
The summation over $\varepsilon_{n^{\,\prime}}$ ($\not=\varepsilon_{n})$ in Eq.\ (\ref{eq:2dG01})  
is performed in accordance with the 3D case, and, instead of }
{Eq.\ (\ref{eq:G6}), we obtain 
\begin{eqnarray}
& &
\Gamma_{\nu}\ks{\mathbf{k},\rm{i}\varepsilon_{n}}
\simeq 
{\bar \Gamma}_{\nu}\ks{\mathbf{k},\rm{i}\varepsilon_{n}}
+J\gamma_{\nu}({\bf k}-{\bf Q})\,\int_{0}^{2\pi}\frac{{\rm d}\varphi}{2\pi}
\left[
\int_{\pi T}^{\infty}{\rm d}\varepsilon^{\prime\prime}
\frac{1}{\sqrt{a+C^{*}\varepsilon^{\prime\prime}}
(\sqrt{a+C^{*}\varepsilon^{\prime\prime}}+\varepsilon^{\prime\prime})^{2}}
\right.
\nonumber
\\
& &
\left.
\qquad\qquad\qquad\qquad\qquad\qquad\qquad\qquad\qquad
+\pi T\,\frac{2}{a^{2}}\varepsilon_{n}+
2{\rm i}b\left(-\frac{\pi T}{a^{2}}+\frac{\varepsilon_{n}}{a^{2}}\right)\right]
+{\cal O}(\varepsilon_{n}^{2}),
\label{eq:G62d}
\end{eqnarray}
}
{
where $a\equiv {\tilde \eta}+K(1-\cos\,\varphi)$ and 
$b\equiv \sqrt{K/2}(1-\cos\,\varphi)$.  }
{
Then, instead of Eq.\ (\ref{eq:G6b}), the linear terms of 
$\Gamma_{\nu}\ks{\mathbf{k},\rm{i}\varepsilon_{n}}$ in $\varepsilon_{n}$ 
is obtained as follows:
\begin{eqnarray}
& &
\Gamma_{\nu}\ks{\mathbf{k},\rm{i}\varepsilon_{n}}
-\Gamma_{\nu}\ks{\mathbf{k},\rm{i}\varepsilon_{n}}|_{\varepsilon_{n}=0}
\simeq 
{\bar \Gamma}_{\nu}\ks{\mathbf{k},\rm{i}\varepsilon_{n}}
-{\bar \Gamma}_{\nu}\ks{\mathbf{k},\rm{i}\varepsilon_{n}}|_{\varepsilon_{n}=0}
\nonumber
\\
& &
\qquad\qquad\qquad\qquad\qquad\qquad\qquad
+J\gamma_{\nu}({\bf k}-{\bf Q})\,\int_{0}^{2\pi}\frac{{\rm d}\varphi}{2\pi}
\left[
\pi T\,\frac{2}{a^{2}}\varepsilon_{n}+2\frac{b}{a^{2}}({\rm i}\varepsilon_{n})
\right].
\label{eq:G6b2d}
\end{eqnarray}
}

{Finally, substituting Eqs.\ (\ref{eq:2dG012}), (\ref{eq:2dG014}), and (\ref{eq:2dG8}) 
into Eq.\ (\ref{eq:G6b2d}), we obtain
}
{
\begin{equation}
\Gamma_{\nu}\ks{\mathbf{k},\rm{i}\varepsilon_{n}}
\simeq
\Gamma_{\nu}\ks{\mathbf{k},\rm{i}\varepsilon_{n}}|_{\varepsilon_{n}=0}
+J\gamma_{\nu}({\bf k}-{\bf Q})\left[
-\frac{Ak_{\rm F}^{2}}{(k_{\rm F}v_{\rm F})^{4}}\frac{\pi T}{\eta}
+\frac{\sqrt{Ak_{\rm F}^{2}}}{4(k_{\rm F}v_{\rm F})^{3}}\frac{1}{\sqrt{\eta}}
\right]({\rm i}\varepsilon_{n})+{\cal O}(\varepsilon_{n}^{2}).
\label{eq:2dG9}
\end{equation}
Here, we note that the term obtained from the first term in the square bracket of 
Eq.\ (\ref{eq:G6b2d}) cancels out 
the second term in the bracket of Eq.\ (\ref{eq:2dG014}), 
so that the term proportional to $\varepsilon_{n}$ disappears but only that proportional to 
${\rm i}\varepsilon_{n}$ remains. 
As in the case of 3D-AFQCP, this result verifies the fact that the first line in the brace of 
Eq.\ (\ref{eq:G01}), i.e., its real part, is an even function in $\varepsilon_{n}$.   
Equation\ (\ref{eq:2dG9}) gives Eq.\ (46) in the text.  

As in the case of 3d-AFQCP, the term $-2{\rm i}b\,\pi T/a^{2}$ 
in Eq.\ (\ref{eq:G62d}) cancels out the third term in Eq.\ (\ref{eq:2dG013}) or (\ref{eq:2dG014}).
}

{
\section{Derivation of Eq.\ (46) in the text for 2D-FQCP}
Here, we derive Eq.\ (46) in the text for the case of 2D-FQCP. We start with 
Eq.\ (\ref{eq:2dG01}), 
the same equation as that in the case of 2D-AFQCP except that $C^{*}$, in the parameter 
${\tilde a}_{\pm}\equiv a+C^{*}|\pm\varepsilon_{n}-\varepsilon_{n^{\,\prime}}|$, is not constant 
but $C^{*}\equiv {\tilde C}^{*}/\sqrt{1-\cos\,\varphi}$ with ${\tilde C}^{*}$ being a constant. 
Then, we  obtain the same expression for ${\bar \Gamma}_{\nu}\ks{\mathbf{k},\rm{i}\varepsilon_{n}}$ 
as Eq.\ (\ref{eq:2dG02}).  
}
{However, since Eq.\ (\ref{eq:2dG02}) does not include $C^{*}$, the expression 
Eq.\ (\ref{eq:2dG04}) for ${\bar \Gamma}_{\nu}\ks{\mathbf{k},\rm{i}\varepsilon_{n}}$ 
can be applied to the present case of 2D-FQCP:  
\begin{eqnarray}
& &
{\bar \Gamma}_{\nu}\ks{\mathbf{k},\rm{i}\varepsilon_{n}}
=J\gamma_{\nu}({\bf k})\,\pi T\,
\,\int_{0}^{2\pi}\frac{{\rm d}\varphi}{2\pi}
\left\{\frac{1}{[{\tilde \eta}+K(1-\cos\,\varphi)]^{3/2}}-
\frac{2\varepsilon_{n}}{[{\tilde \eta}+K(1-\cos\,\varphi)]^{2}}
\right.
\nonumber
\\
& &
\left.
\qquad\qquad\qquad\qquad
+2{\rm i}\sqrt{\frac{K}{2}}\frac{(1-\cos\,\varphi)}
{[{\tilde \eta}+K(1-\cos\,\varphi)]^{2}}
-6{\rm i}\sqrt{\frac{K}{2}}\frac{\varepsilon_{n}}
{[{\tilde \eta}+K(1-\cos\,\varphi)]^{5/2}}
\right\}.
\label{eq:G042dF}
\end{eqnarray}
Then, the expression Eq.\ (\ref{eq:2dG014}) for 
${\bar \Gamma}_{\nu}\ks{\mathbf{k},\rm{i}\varepsilon_{n}}$ is also valid in the case of 
2D-FQCP:
\begin{equation}
{\bar \Gamma}_{\nu}\ks{\mathbf{k},\rm{i}\varepsilon_{n}}
\simeq
J\gamma_{\nu}({\bf k}-{\bf Q})\,
\left[
\frac{Ak_{\rm F}^{2}}{(k_{\rm F}v_{\rm F})^{3}}\frac{T}{\eta}
-\,\varepsilon_{n}\,\frac{(Ak_{\rm F}^{2})^{3/2}}{2(k_{\rm F}v_{\rm F})^{4}}
\frac{\pi T}{\eta^{3/2}}
+{\rm i}\frac{\sqrt{Ak_{\rm F}^{2}}}{4(k_{\rm F}v_{\rm F})^{3}}
\frac{\pi T}{\sqrt{\eta}}
-{\rm i}\varepsilon_{n}\,\frac{Ak_{\rm F}^{2}}{(k_{\rm F}v_{\rm F})^{4}}
\frac{T}{\eta}
\right].
\label{eq:2dG014F}
\end{equation}
Since $\eta\propto (-T\,\log\,T)$ in the case of 2D-FQCP,~\cite{Hatatani} 
the first and the fourth terms in Eq.\ (\ref{eq:2dG014}) vanish, in the limit $T\to 0$, 
as $1/(-\log\,T)$, while the second term diverges as $1/[\sqrt{T}(-\log\,T)^{3/2}]$. 
On the other hand, the first term in Eq.\ (\ref{eq:2dG014F}) vanishes 
as $1/(-\log\,T)$ in the limit $T\to 0$, which should be included in Eq.\ (44) in the text 
because it gives a shaper cusp in $T$ dependence in the limit $T\to 0$ 
than the second term of Eq.\ (44) in the text.  

The summation over $\varepsilon_{n^{\,\prime}}$ ($\not=\varepsilon_{n})$ in Eq.\ (\ref{eq:2dG01})  
is performed in the same way as the case of 2D-AFQCP, and we obtain the same relation as 
Eq.\ (\ref{eq:G6b2d}):  
\begin{eqnarray}
& &
\Gamma_{\nu}\ks{\mathbf{k},\rm{i}\varepsilon_{n}}
-\Gamma_{\nu}\ks{\mathbf{k},\rm{i}\varepsilon_{n}}|_{\varepsilon_{n}=0}
\simeq 
{\bar \Gamma}_{\nu}\ks{\mathbf{k},\rm{i}\varepsilon_{n}}
-{\bar \Gamma}_{\nu}\ks{\mathbf{k},\rm{i}\varepsilon_{n}}|_{\varepsilon_{n}=0}
\nonumber
\\
& &
\qquad\qquad\qquad\qquad\qquad\qquad\qquad
+J\gamma_{\nu}({\bf k}-{\bf Q})\,\int_{0}^{2\pi}\frac{{\rm d}\varphi}{2\pi}
\left[
\pi T\,\frac{2}{a^{2}}\varepsilon_{n}+2\frac{b}{a^{2}}({\rm i}\varepsilon_{n})
\right].
\label{eq:2dG7F}
\end{eqnarray}
where $a\equiv {\tilde \eta}+K(1-\cos\,\varphi)$ and 
$b\equiv \sqrt{K/2}(1-\cos\,\varphi)$.  
Then, $\Gamma_{\nu}\ks{\mathbf{k},\rm{i}\varepsilon_{n}}$ is given by the same expression as 
Eq.\ (\ref{eq:2dG9}): 
\begin{equation}
\Gamma_{\nu}\ks{\mathbf{k},\rm{i}\varepsilon_{n}}
\simeq
\Gamma_{\nu}\ks{\mathbf{k},\rm{i}\varepsilon_{n}}|_{\varepsilon_{n}=0}
+J\gamma_{\nu}({\bf k}-{\bf Q})\left[
-\frac{Ak_{\rm F}^{2}}{(k_{\rm F}v_{\rm F})^{4}}\frac{\pi T}{\eta}
+\frac{\sqrt{Ak_{\rm F}^{2}}}{4(k_{\rm F}v_{\rm F})^{3}}\frac{1}{\sqrt{\eta}}
\right]({\rm i}\varepsilon_{n})+{\cal O}(\varepsilon_{n}^{2}).
\label{eq:2dG9F}
\end{equation}
This gives Eq.\ (46) in the text.  

As in the case of 2d-AFQCP, the term $-2{\rm i}b\,\pi T/a^{2}$ 
in Eq.\ (\ref{eq:G62d}) cancels out the third term in Eq.\ (\ref{eq:2dG014F}).
}